\documentclass[epj]{svjour}

\usepackage{amssymb}
\usepackage{amsmath}
\usepackage{graphicx}
\usepackage{pifont}
\usepackage{supertabular}
\usepackage{dcolumn}

\begin{document}

%%%%%%%%%%%%%%%%%%%%%%%%%%%%%%%%%%%%%%%%%%%%%%%%%%%%%%%%%%%%%%%%%%%
%%%
%%%     T I T E L
%%%
\title{Compton Scattering by the Proton using a Large-Acceptance 
       Arrangement\thanks{Supported
       by the Italian Istituto Nazionale di Fisica Nucleare (INFN) and by
       Deutsche Forschungsgemeinschaft (SFB 201) and by DFG-contracts
       Schu222 and 436RUS113/510}}
%\subtitle{Do you have a subtitle?\\ If so, write it here}
\author{S. Wolf\inst{1}, V. Lisin\inst{2}, R. Kondratiev\inst{2},
        A.M. Massone\inst{3}, G. Galler\inst{1}, J. Ahrens\inst{4},
        H.-J. Arends\inst{4}, R. Beck\inst{4}, M. Camen\inst{1},
        G.~P.~Capitani\inst{5}, P. Grabmayr\inst{6},
        F. H\"arter\inst{4}, T. Hehl\inst{6}, P. Jennewein\inst{4},
        K. Kossert\inst{1}, A.I. L'vov\inst{7}, C.~Molinari\inst{3},
        P. Ottonello\inst{3}, R.O. Owens\inst{8},  J. Peise\inst{4}, 
        I. Preobrajenskij\inst{4},
        S. Proff\inst{1}, A. Robbiano\inst{3}, M. Sanzone\inst{3},
        M. Schumacher\inst{1}\thanks{\emph{Corresponding author:}
        schumacher@physik2.uni-goettingen.de; 
        http://www.physik.uni-goettingen.de/members/Schumacher.Martin.html}, 
        M. Schmitz\inst{4},
        F.~Wissmann\inst{1}}
%\offprints{}          % Insert a name or remove this line
\institute{Zweites Physikalisches Institut, Universit\"at G\"ottingen,
           D-37073 G\"ottingen, Germany \and
           Institute for Nuclear Research, Moscow 117312, Russia \and
           Dipartimento di Fisica dell'Universita di Genova and
           INFN - Sezione di Genova, I-16146 Genova, Italy \and
           Institut f\"ur Kernphysik, Universit\"at Mainz, D-55099 Mainz,
           Germany \and
           INFN - Laboratori Nazionali di Frascati, I-00044 Frascati,
           Italy \and
           Physikalisches Institut, Universit\"at T\"ubingen, D-72076
           T\"ubingen, Germany \and
           P. N. Lebedev Physical Institute, Moscow 117924, Russia\and
            Kelvin Laboratory, Glasgow University, Glasgow, Scotland,
           G12 8QQ,UK}
\date{Received: date / Revised version: date}
% The correct dates will be entered by Springer

%%%%%%%%%%%%%%%%%%%%%%%%%%%%%%%%%%%%%%%%%%%%%%%%%%%%%%%%%%%%%%%%%%%
%%%
%%%     A B S T R A C T
%%%
\abstract{Compton scattering by the proton has been measured over a wide
          range covering photon energies $250~\mathrm{MeV} \lesssim
          E_\gamma \lesssim 800~~\mathrm{MeV}$ and photon scattering angles
          $30^\circ \lesssim \theta^\mathrm{lab}_\gamma  \lesssim 150^\circ$,
          using the tagged-photon facility at MAMI (Mainz) and the
          large-acceptance arrangement LARA. The previously existing data
          base on proton Compton scattering is greatly enlarged by more
          than 700 new data points. The new data are interpreted in terms
          of dispersion theory based on the SAID-SM99K parametrization of
          photo-meson amplitudes. It is found that two-pion exchange in the
          $t$-channel is needed for a
          description of the data in the second resonance region. 
          The data are well represented if this channel is modeled
          by a single pole with the mass parameter $m_\sigma \approx$ 600 MeV.
          The
          asymptotic part of the spin dependent amplitude is found to be well
          represented by $\pi^0$-exchange in the $t$-channel. No indications
          of additional effects were found.
         Using the mass parameter $m_\sigma$ of the two-pion exchange
          determined from the second resonance region and using the
         new global average for the difference of the electric and magnetic
          polarizabilities of the proton, $\alpha - \beta = \left(10.5
          \pm 0.9_\mathrm{stat+syst}\pm 0.7_\mathrm{model} \right) 
         \times 10^{-4}~\mathrm{fm}^3$, as
          obtained from a recent  experiment on proton Compton scattering
          below pion photoproduction threshold, a backward spin-polarizability
          of $\gamma_{\pi} = \left( - 37.1 \pm 
          0.6_\mathrm{stat+syst} \pm 3.0_\mathrm{model} \right)
          \times 10^{-4}~\mathrm{fm}^4$ has been determined from data
          of the first resonance region below 455 MeV. This value is in a good
          agreement with predictions of dispersion relations and chiral
          perturbation theory. 
          From a subset of data between 280 and 360 MeV the resonance
          pion-photoproduction amplitudes were evaluated leading to a E2/M1
          multipole ratio of the 
          p $\to$ $\Delta$ radiative transition of 
          EMR(340 MeV)= $(-1.7 \pm 0.4_{\rm stat+syst}\pm 0.2_{\rm
          model})$\%. It was found that this number is dependent  on the
          parameterization of photo-meson amplitudes. With the MAID2K 
          parameterization an  E2/M1 multipole ratio of
          EMR(340 MeV)= $(-2.0 \pm 0.4_{\rm stat+syst}\pm 0.2_{\rm
          model})$\% is obtained.
\PACS{
      {25.20.Dc}{Compton scattering, spin polarizabilities, proton,
                 scattering amplitudes}%   \and
%      {PACS-key}{discribing text of that key}
     } % end of PACS codes
} %end of abstract

%%%%%%%%%%%%%%%%%%%%%%%%%%%%%%%%%%%%%%%%%%%%%%%%%%%%%%%%%%%%%%%%%%%
%%%
%%%     T I T L E P A G E
%%%
\authorrunning{S. Wolf et al.}
\titlerunning{Compton Scattering by the Proton}
\maketitle

%%%%%%%%%%%%%%%%%%%%%%%%%%%%%%%%%%%%%%%%%%%%%%%%%%%%%%%%%%%%%%%%%%%
%%%
%%%     I N T R D U C T I O N
%%%
\section{Introduction}
\label{intro}

Elastic scattering of photons from the proton (proton Compton scattering) 
is known \cite{baranov76,petrunkin81} to be a valuable tool
for investigations of the structure of the nucleon. 
The specific feature of this process is that it depends on  
the electromagnetic interaction only and, therefore, is
especially suited to study electromagnetic properties of the nucleon.
Nevertheless, it took a long time until decent use could be made of the
method. One reason for the delay was that the process is difficult to measure
and, therefore, the data base remained fragmentary.  The other reason 
was that the methods of data interpretation were not well enough 
developed, so that definite conclusions on the electromagnetic 
properties of the nucleon could
not be drawn with the desired precision. The present work 
shows that by now the shortcomings of the previous approaches have been
overcome due to new experimental techniques applied here for the first time
in a Compton scattering experiment
and due to recent and continuing progress in developing the dispersion theory
of Compton scattering.

The properties of the nucleon accessible by a given experiment depend on the
type of the reaction. In Compton scattering properties are selected which are
specific for two-photon interactions. These are the electromagnetic
polarizabilities and spin polarizabilities in  first place and specific
$t$-channel exchanges. Furthermore, due to the optical theorem and dispersion
relations there is a close relation to meson photoproduction. This implies
that Compton scattering also is a good tool of nucleon spectroscopy for
measurements of  strengths and multipolarities of electromagnetic 
transitions.

An exhaustive review of literature on proton Compton scattering in the energy
region of nucleon resonances up to 1974 has been published by Baranov and
Fil'kov \cite{baranov76}. Shortly thereafter experiments have been carried out
in Bonn \cite{genzel76,jung81} and Tokyo \cite{ishii80,wada84} which led to
essential progress.  The main difficulty in measuring Compton scattering
by the proton above the meson photoproduction threshold consists in the
separation of the ($\gamma$,$\gamma$) from the ($\gamma$,$\pi^0$) reaction
channel.  This difficulty has led to different strategies depending on the
available photon facility and the detection system. Because of the
absence of high duty-factor electron beams and connected with that, the
absence of high fluxes of tagged photons, the previous experiments had to be
carried out with bremsstrahlung
\cite{genzel76,jung81,ishii80,wada84}. As long as the experiments were
restricted to the $\Delta$ energy range \cite{genzel76} scintillator
telescopes for the proton and the detection of the shower produced by the
photon were sufficient. At higher energies \cite{jung81,ishii80,wada84} the
lack of information on the energy of the primary photon had to be compensated
by high-resolution proton spectrometry which required the use of magnetic
spectrometers in combination with high angular-resolution track reconstruction.
By achieving also a good position resolution of the photon it was then 
possible to
measure $p-\gamma$ directional correlations with high angular resolution.  In
the Bonn set-up \cite{jung81} a large-volume NaI(Tl) detector was used 
with photomultipliers on the front side to locate the incidence
point of the photon.  In the Tokyo set-up \cite{ishii80,wada84} a lead glass
\v{C}erenkov counter was used in combination with a lead plate $\gamma \to
e^+e^-$ converter and two multi-wire proportional chambers.  
When applying this method, the events from the two reaction channels
$(\gamma,\gamma)$ and $(\gamma,\pi^0)$ differ in the widths of the 
$p-\gamma$ angular correlations. Therefore, the $(\gamma,\gamma)$ events
show up as a narrow peak on top of a broad background. Though this method
leads to a comparatively safe separation of events, it has the disadvantage
that one setting of the apparatus leads to only one differential cross section
per given angular and energy interval.

At modern facilities with tagged photons experiments providing
only one differential cross section per given angular and energy interval are
not in line with the required economic use of  the beam. When using tagged
photons together  with a large-volume NaI(Tl) detector it is relatively
easy to separate the two types of events through the good energy resolution of
the NaI(Tl) detector over the whole energy range of the $\Delta$ resonance.
This method has been applied in Compton scattering experiments
by the proton carried out at the tagged-photon facilities at Saskatoon (SAL)
\cite{hallin93}, Brook\-haven (LEGS) \cite{blanpied96,blanpied97,tonnison98} 
and Mainz (MAMI) \cite{peise96,huenger97,wissmann99}. The advantage of
this method is that the recoil proton has not necessarily to be detected, so
that there is no restriction in the accessibility of small photon angles and
low photon energies, where the recoil proton does not leave the target with
sufficiently high energy to reach the detector. The disadvantages 
are the restriction to the $\Delta$ energy range and the accessibility 
of only one scattering angle per
beam-time period. In an other experiment carried out at MAMI (Mainz)
\cite{molinari96} the apparatus determined the full set of kinematical
variables of the photon and the proton. The protons were detected
using an  $E$--$\Delta E$ plastic scintillator telescope the photons were 
registered by lead glass detectors.

The LARA ({\bf LAR}ge {\bf A}cceptance) experiment is the 
first Compton scattering experiment where the restrictions discussed above 
were overcome and a large angular range from 
$\theta^\mathrm{lab}_\gamma=30^\circ$ to $150^\circ$ and large energy 
range from E$_\gamma$ = 250 MeV to 800 MeV is covered
simultaneously with one experimental set-up. This is achieved by the use
of the tagging method in combination with large acceptance arrangements for
the recoil proton and the scattered photon. In principle, 
the apparatus determines the full set
of kinematical variables of the proton and the photon and contains many
features of the Bonn \cite{jung81} and Tokyo \cite{ishii80,wada84} designs,
except for the fact that magnetic spectrometers are incompatible with large
angular and energy acceptance detection. Therefore, the proton spectrometry
had to rely on time-of-flight measurements using long flight
paths. Except for the available space, the limitations of this method are
given by the energy loss and the straggling of the protons in air. Due to
straggling the proton angle cannot be determined to much better than $\Delta
\theta_\mathrm{p}=\pm 1^\circ$ corresponding to a photon interval of $\Delta
\theta_\gamma=\pm 2^\circ$ for the Compton kinematics.  This was the
underlying point of view when selecting the angular resolutions for the
photons and protons in the apparatus design.  The expected properties of the
LARA experimental set-up have been explored in detailed simulation studies
\cite{falkenberg95}.  In these studies it was shown that by combining
$p-\gamma$ angular correlation with time of flight measurements an event by
event separation of $(\gamma,\gamma)$ and $(\gamma,\pi^0)$ events should be
possible in the energy region of the first resonance and that this property
should be partly preserved in the second resonance region.

The dispersion theory of Compton scattering by the nucleon which formerly was
restricted to the first resonance
\cite{baranov76,pfeil74,guiasu78,akhmedov81,lvov81,lvov85} has been extended
to cover also the second resonance region \cite{lvov97}. This dispersion
theory proved to be much more precise than alternative approaches based on a
phenomenological resonance model \cite{ishii80,wada84}, where the scattering
amplitude is represented as a sum of Breit-Wigner nucleon resonances and an
adjusted real background which is assumed to be a modified Born term.  Even
after the development of improved resonance models, in which a K-matrix
unitarization is implemented \cite{hida76,feuster99,korchin98,korchin00}, the
dispersion theory still provides the highest precision.

The quantitative success of the dispersion theory supports the expectation
that Compton scattering may be used as a precise tool for measuring several
electromagnetic properties of the nucleon, including in particular the
electric and magnetic polarizabilities $\alpha$ and $\beta$, the four
so-called spin polarizabilities $\gamma_i$ (the backward spin polarizability
$\gamma_\pi$ being a particular linear combination of them), the strength
$M_{1+}$ and the multipole ratio E2/M1 of the $N\to\Delta$ transition.  These
quantities enter into the theoretical Compton differential cross section as
(not fully independent) parameters and they are predominantly important in the
$\Delta$ energy range.

The dispersion theory described in \cite{lvov97} has recently been improved in
some aspects by Drechsel et al.\ \cite{drechsel99} using subtracted dispersion
relations. The main difference of the recent version \cite{drechsel99}
compared to the former one \cite{lvov97} is that, like in
\cite{guiasu78,akhmedov81} and some older works, the two-pion $t$-channel
exchange was implemented in an explicit way in order to fix otherwise
uncertain so-called asymptotic contributions to the invariant amplitudes $A_1$
and $A_2$.  Theoretically such an improvement is very important because it has
the potential to remove free parameters which are specific for nucleon Compton
scattering.  Practically, however, free parameters do not disappear
completely, since the $t$-channel exchanges are not exhausted by low-lying
$\pi^0$ and $\pi\pi$ states.  Thus, a poorly-known input from high-energy
contributions actually remains in the theory.

The difference between the two versions of the dispersion theory have been
found by us to be small in the $\Delta$ energy range but still have to be
explored for higher energies where at present no predictions from subtracted
dispersion relations are available.

The dispersion theory in the version of L'vov et al.\ \cite{lvov97} utilizes a
less sophisticated  phenomenological approach for a description of $t$-channel
exchanges. In the formalism of unsubtracted fixed-$t$ dispersion relations
used there, these are the asymptotic contributions $A_i^{\rm as}(\nu,t)$ which
car\-ry the information on $t$-channel exchanges.  These contributions are
theoretically expected to be energy independent at energies $E_\gamma$ well
below the cutoff $E_\gamma^\mathrm{max} \approx 1.5$ GeV used for separating
the asymptotic region. Therefore, only a $t$-dependence of $A_i^{\rm
  as}(\nu,t)$ is taken into account. Practically, these amplitudes are
parameterized by pole $t$-channel exchanges associated with the lightest
mesons. In particular, the asymptotic contribution $A_1^{\rm as}$ is
parameterized by an effective $\sigma$-exchange which therefore introduces an
adjustable parameter, $m_\sigma$, which can be loosely interpreted as a
(effective) mass of the $\sigma$ meson. The product of couplings of the
$\sigma$ meson to the photon and the nucleon constitutes 
one additional parameter, which is
fixed using an experimental number for the difference, $\alpha-\beta$,
of the electric and magnetic polarizabilities of the proton.

One other large asymptotic contribution, $A_2^{\rm as}$, is assumed to be
given by $\pi^0$-exchange. An important question raised by Tonnison et
al.\ \cite{tonnison98} is, whether the $\pi^0$-exchange indeed exhausts the
asymptotic contribution to the amplitude $A_2$ or, alternatively, an
additional large background exists in $t$-channel exchanges with the quantum
numbers of pseudoscalar mesons.  In the latter case, such a background can
largely modify the backward spin polarizability of the proton, $\gamma_\pi$,
which therefore becomes an important signature of the $t$-channel dynamics of
Compton scattering.

Another feature of the theory \cite{lvov97} is that it takes into account an
important channel of double-pion photoproduction including
$\pi\Delta$-production. In  forward direction the contribution of the
$2\pi$-channel to the Compton scattering amplitude is well known. Its
extension to nonforward angles requires a further consideration of the
multipole structure of double-pion photoproduction (see \cite{lvov97} for
details). The result may then be tested by experiments carried out in the
second resonance region.

The present paper contains an exhaustive description of the results
of the LARA experiment and their interpretation in terms of the currently
accepted dispersion theory \cite{lvov97} based on the SAID-SM99K 
\cite{arndt96} multipole
analysis and specific models to take into account  asymptotic 
contributions or subtractions. For comparison also the MAID2K \cite{maid}
parameterization has been applied.
A short version of this work
has been published elsewhere \cite{galler01}.

In contrast to our present approach, the realistic $2\pi$-exchange in the
$t$-channel does not correspond to a narrow resonance but rather to a broad
continuum. This apparent deficiency of our approach
does not show up as a discrepancy when
comparing the present experimental data with predictions. However, from a
theoretical point of view this deficiency is not acceptable and should be
removed.  This will be done in a following paper which is devoted to
improvements of the dispersion theory and to further interpretations.

%%%%%%%%%%%%%%%%%%%%%%%%%%%%%%%%%%%%%%%%%%%%%%%%%%%%%%%%%%%%%%%%%%%
%%%
%%%     S E C T I O N       Experiment
%%%
\section{Experiment}
\label{sec:experiment}
\begin{figure}[t]
  \centering \includegraphics[width=0.85\columnwidth]{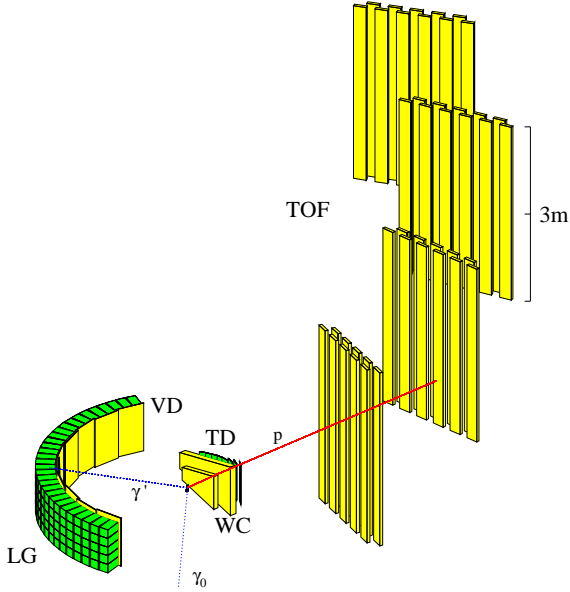}
  \caption{Perspective view of the LARA arrangement. Left: Photon arm
           consisting of 10 blocks \`a 3 $\times$ 5 lead glass detectors
           (LG) each block equipped with a 1~cm plastic scintillator (VD).
           Right: Proton arm  consisting of two wire chambers (WC) at
           distances of 25 and 50 cm from the target center, 8 plastic
           scintillators serving as trigger detectors (TD) and 43 bars of
           20 cm $\times$ 300 cm $\times$ 5 cm plastic scintillators serving
           as time-of-flight (TOF) stop detectors. The scattering target
           consisted of lq. $\mathrm{H}_2$ contained in a 3 cm
           $\varnothing$ $\times$ 20 cm Kapton cylinder.}
  \label{fig:fig1}
\end{figure}
\begin{figure}[t]
  \centering \includegraphics[width=0.85\columnwidth]{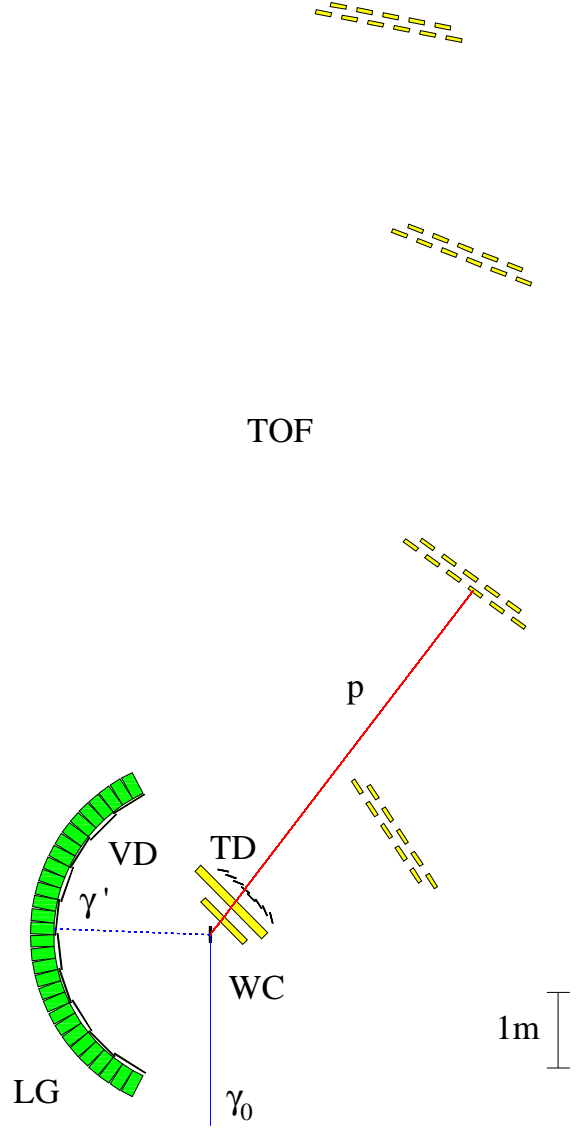}
  \caption{Same as Figure \ref{fig:fig1} but projected into the horizontal
           plane}
  \label{fig:fig2}
\end{figure}
\begin{figure}[t]
  \centering \includegraphics[width=0.85\columnwidth]{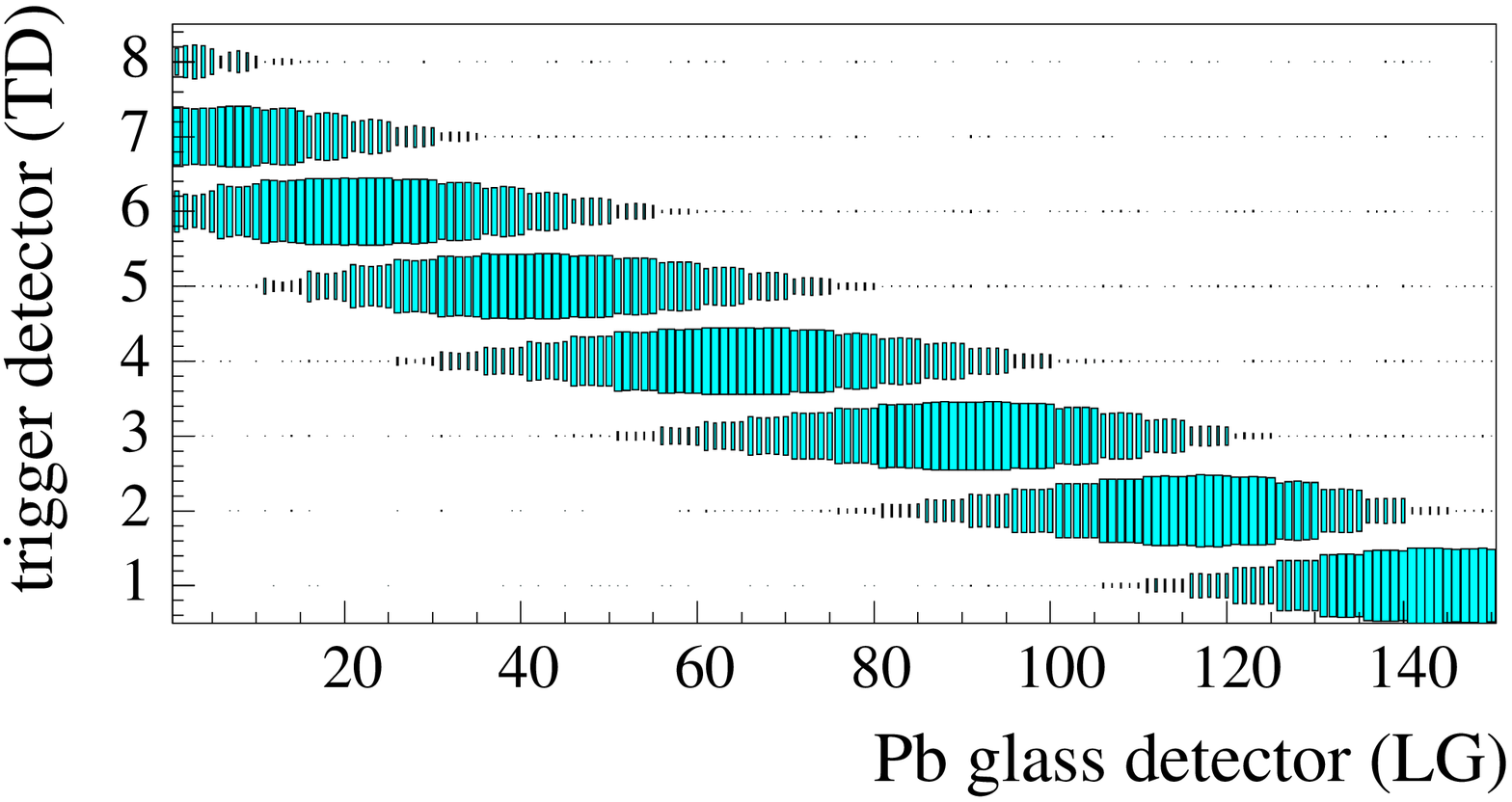}
  \caption{Simulated scatter plot of events showing the correlation between
           trigger detector (TD) and Pb glass detectors (LG) for Compton
           events. Abscissa: Number of the Pb glass detector with the
           detectors 1--5 covering the angular range from
           $\theta^\mathrm{lab}_\gamma = 25^\circ$ to $30^\circ$ and the
           detectors 146--150 covering the angular range from 
           $\theta^\mathrm{lab}_\gamma = 150^\circ$ to $155^\circ$. The trigger
           detector 1 is closest to the forward direction and covers the
           range of proton angles from
           $\theta^\mathrm{lab}_\mathrm{p} = 7.3^\circ$ to $15.6^\circ$. The
           correlation between the two angles $\theta^\mathrm{lab}_\gamma$ and
           $\theta^\mathrm{lab}_\mathrm{p}$ was used to preselect Compton
           events through the trigger condition.}
  \label{fig:fig3}
\end{figure}
The present paper contains the results of an experiment carried out using the
{\bf LAR}ge {\bf A}cceptance arrangement (LARA) shown in Fig.~\ref{fig:fig1}
as a perspective view from the side. The same apparatus is shown 
in Fig.~\ref{fig:fig2} as viewed
from the top.  This arrangement was designed to cover the  angular range
of photon scattering-angles from 
$\theta^\mathrm{lab}_\gamma = 30^\circ$ to $150^\circ$ in
the laboratory and the interval of photon energies from 
E$_\gamma$ = $250 \; \mathrm{MeV}$
to $800 \; \mathrm{MeV}$ with limitations given by the range of protons in the
scattering target. Due to the energy loss in the scattering 
target the minimum energy of a proton to be detected is 
about 30 MeV. This leads to the unwanted restriction that the small-angle
low-energy section of the photon range given above 
is not accessible.  However, this
range should easily be accessible by an experiment with a large-volume NaI(Tl)
detector like the Mainz 48 cm $\varnothing$ $\times$ 64 cm NaI(Tl) detector
\cite{huenger97}. This detector has  sufficient energy resolution in this
range to separate photons from the $(\gamma,\gamma)$ and $(\gamma,\pi^0)$
reactions so that the recoil protons have not to be detected.

The experiment makes use of the tagged photon facility \cite{anthony91}
installed at the $855 \; \mathrm{MeV}$ three-stage microtron MAMI in Mainz
\cite{herminghaus76}. The energy resolution achieved by the tagger was $\Delta
E_\gamma = 2 \; \mathrm{MeV}$ on the average. The maximum rate of tagged
photons as limited by the tagger is $10^5 \; \mathrm{s}^{-1}$ per tagger
channel. In the present case this rate was lower by a factor of about two
because of limitations due to the wire chambers.

The scattering target consists of lq.~$\mathrm{H}_2$ contained in a Kapton
cylinder of $200 \; \mathrm{mm}$ length and $30 \; \mathrm{mm}$ diameter. The
apparatus (Figs. 1 and 2) consists of 150 lead glass 
photon detectors (LG) having dimensions
of $15 \; \mathrm{cm} \times 15 \; \mathrm{cm} \times 30 \; \mathrm{cm}$
positioned cylindrically around the scattering target with the front faces
having distances of $200 \; \mathrm{cm}$ from the target center.  This leads
to an angular resolution on the photon arm  of $\pm 2.2^\circ$ both in the
horizontal and the vertical direction.  Each block containing 3 (horizontal)
$\times$ 5 (vertical) detectors is equipped with a plastic scintillator (VD)
of $1 \; \mathrm{cm}$ thickness to identify charged background.

On the proton arm  of the detector arrangement the proton angle
$\theta_\mathrm{p}$ with respect to the incident photon beam is determined by
two wire chambers (WC) at distances of $25 \; \mathrm{cm}$ and $50 \;
\mathrm{cm}$ from the target center. Each of these wire chambers consists of
two layers of wires tilted against the vertical direction by $+30^\circ$ and
$-30^\circ$, respectively.  The distance between wires in the layers is 2.5
mm.  The resolution achieved for the proton angle is better than $1^\circ$ in
the horizontal (geometrical $0.13^\circ$) and vertical (geometrical
$0.47^\circ$) directions. The time of flight is measured through coincidences
between signals from the tagger and signals from 43 bars of $20 \; \mathrm{cm}
\times 300 \; \mathrm{cm} \times 5 \; \mathrm{cm}$ plastic scintillators (TOF)
\cite{grabmayr98}. The latter are arranged in 4 planes positioned at distances
of 2.6, 5.7, 9.4 and $12.0 \; \mathrm{m}$ from the target center. The
experiment trigger was defined through a coincidence between a signal from a
lead glass block and a signal from one out of 8 trigger detectors (TD)
positioned behind the wire chambers, with the geometry complying with 
the angular
constraints of a Compton event.  The preselection of data possible through the
trigger condition is demonstrated in Fig.~\ref{fig:fig3}, where the
correlation between the trigger detectors and the Pb glass detectors is shown
by a scatter plot of events obtained by computer simulation. Each 5 of the 150
Pb glass detectors are positioned on top of each other so that ranges of 5
successive Pb glass detectors approximately correspond to the same interval of
photon scattering angles.

%%%%%%%%%%%%%%%%%%%%%%%%%%%%%%%%%%%%%%%%%%%%%%%%%%%%%%%%%%%%%%%%%%%
%%%
%%%     S E C T I O N       Data analysis
%%%
\section{Data analysis}
\label{sec:dataanalysis}
\begin{figure}[t]
  \centering \includegraphics[width=0.85\columnwidth]{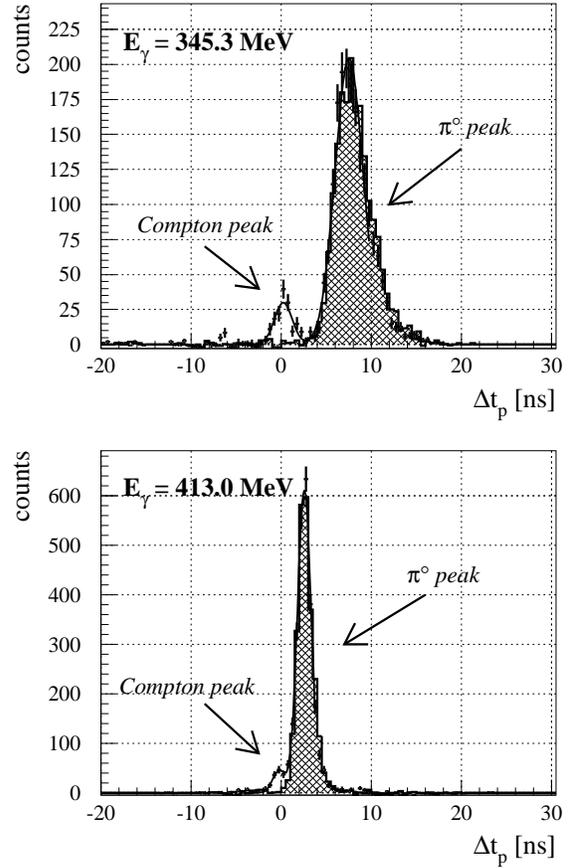}
  \caption{Typical experimental missing time spectra for protons at primary
           photon energies of $E_\gamma = 345.3$ MeV (upper figure) and
           $E_\gamma = 413.0$ MeV (lower figure) measured at a photon angle
           of $\theta^\mathrm{lab}_{\gamma'} = 70^\circ$. The protons were
           detected with one plastic scintillator bar positioned at a proton
           angle of $\theta^\mathrm{lab}_\mathrm{p} = 45^\circ$.}
  \label{fig:fig4}
\end{figure}
\begin{figure}[t]
  \centering \includegraphics[width=0.85\columnwidth]{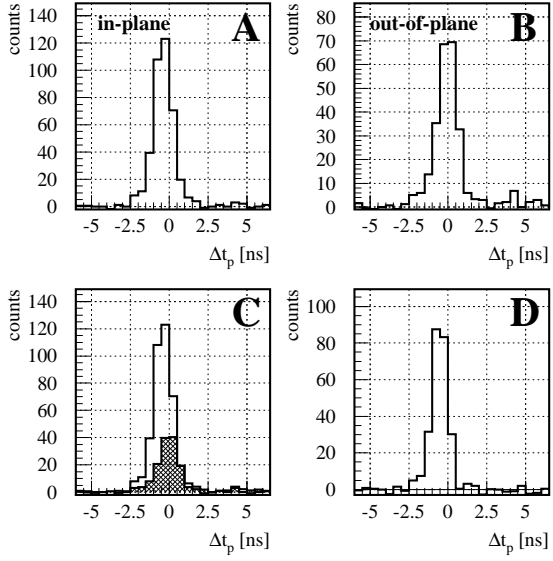}
  \caption{Typical experimental missing time spectrum for protons at a
           primary photon energy of $E_\gamma = 659.3$ MeV measured at a
           photon angle of $\theta^\mathrm{lab}_{\gamma'} = 116^\circ$. The
           protons were detected with 4 plastic scintillator bars positioned
           around a proton angle of
           $\theta^\mathrm{lab}_\mathrm{p} = 20^\circ$. {\bf A}: In-plane data.
           {\bf B}: Out-of-plane data. {\bf C}: In-plane data and the
           corresponding adjusted out-of-plane data (cross-hatched area).
           {\bf D}: Compton events obtained from the in-plane data by
           subtracting the corresponding adjusted out-of-plane data.}
\label{fig:fig5}
\end{figure}
\begin{figure}[t]
  \centering \includegraphics[width=0.85\columnwidth]{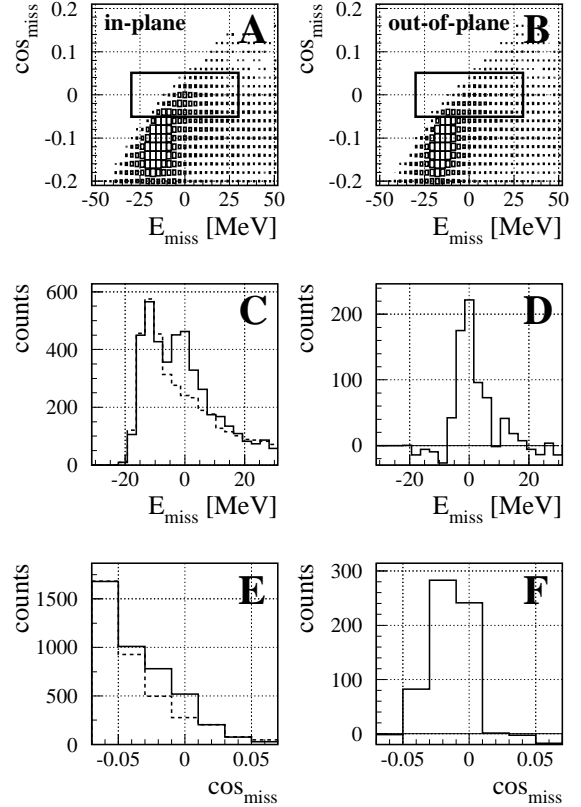}
  \caption{Two-dimensional analysis of experimental data obtained at a photon
           energy of $E_\gamma=467.4$ MeV and a scattering angle of
           $\theta^\mathrm{lab}_{\gamma}=37^\circ$. The data were obtained
           with 4 TOF plastic scintillator bars positioned at proton angles
           around $\theta^\mathrm{lab}_\mathrm{p} = 63^\circ$. {\bf A}: Scatter
           plot of experimental data measured in-plane. {\bf B}: Scatter
           plot of experimental data measured out-of-plane. In these two
           subfigures the rectangular frames denote those ranges where Compton
           events are expected to be located in {\bf A}. {\bf C}: Vertical
           projection of the data inside the rectangular frames, with the
           data from {\bf A} denoted by full lines and the corresponding
           out-of-plane data denoted by a dashed lines. {\bf D}: Same as
           {\bf C} but showing the difference between the solid and the
           dashed lines. {\bf E}: Horizontal projection of the data inside
           the rectangular frames, with the data from {\bf A} denoted by
           full lines and the corresponding out-of-plane data denoted by
           a dashed lines. {\bf F}: Same as {\bf E} but showing the
           difference between the solid and the dashed lines.}
  \label{fig:fig6}
\end{figure}
\begin{figure}[t]
  \centering \includegraphics[width=0.85\columnwidth]{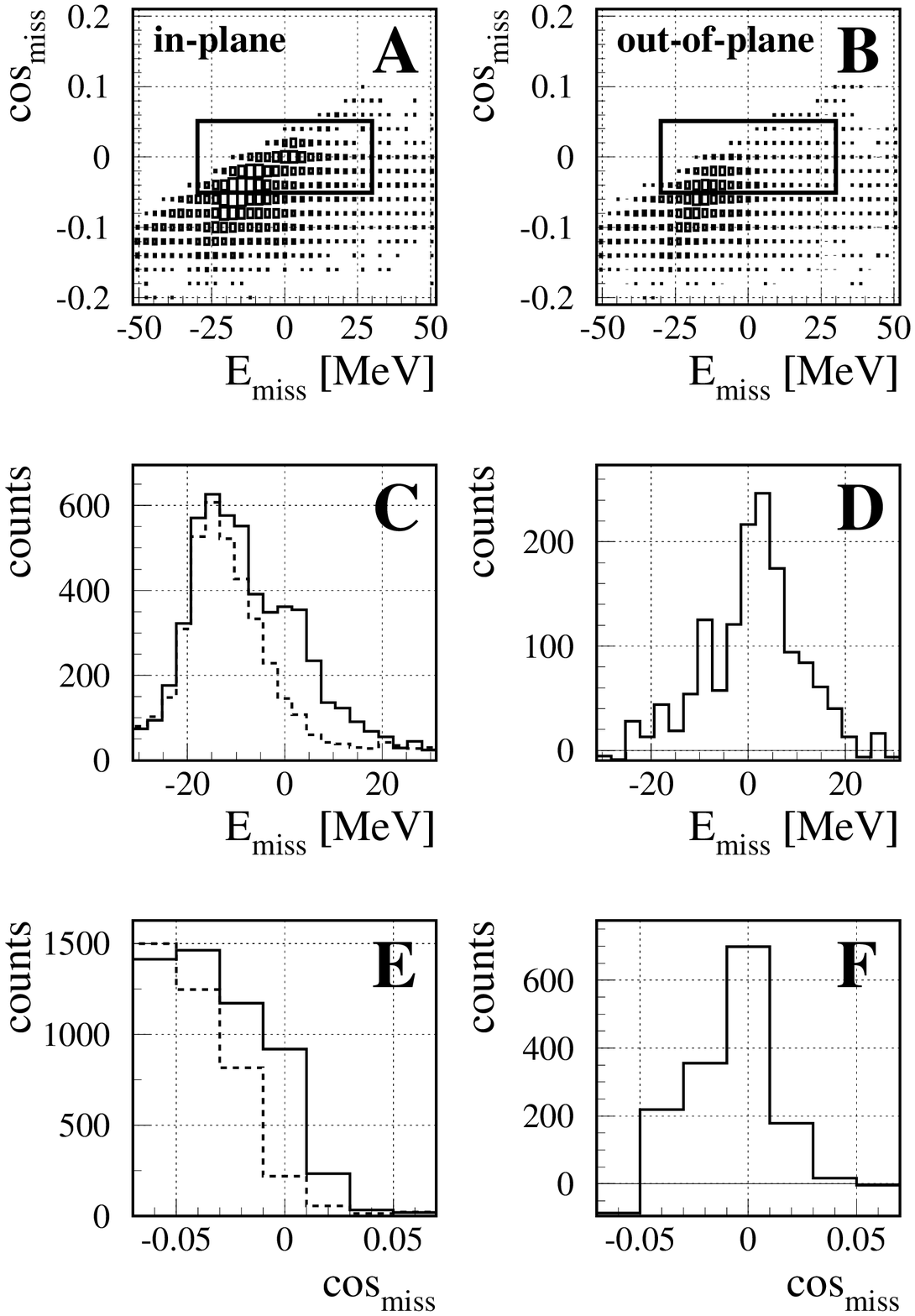}
  \caption{Two-dimensional analysis of experimental data obtained at a photon
           energy $E_\gamma = 657.5$ MeV and a scattering angle of
           $\theta^\mathrm{lab}_{\gamma}=29^\circ$. The data were obtained
           with 4 TOF plastic scintillator bars positioned at proton angles
           around $\theta^\mathrm{lab}_\mathrm{p} = 67^\circ$. For further
           details see Fig.~\ref{fig:fig6}.}
  \label{fig:fig7}
\end{figure}
\begin{figure}[t]
  \centering \includegraphics[width=0.85\columnwidth]{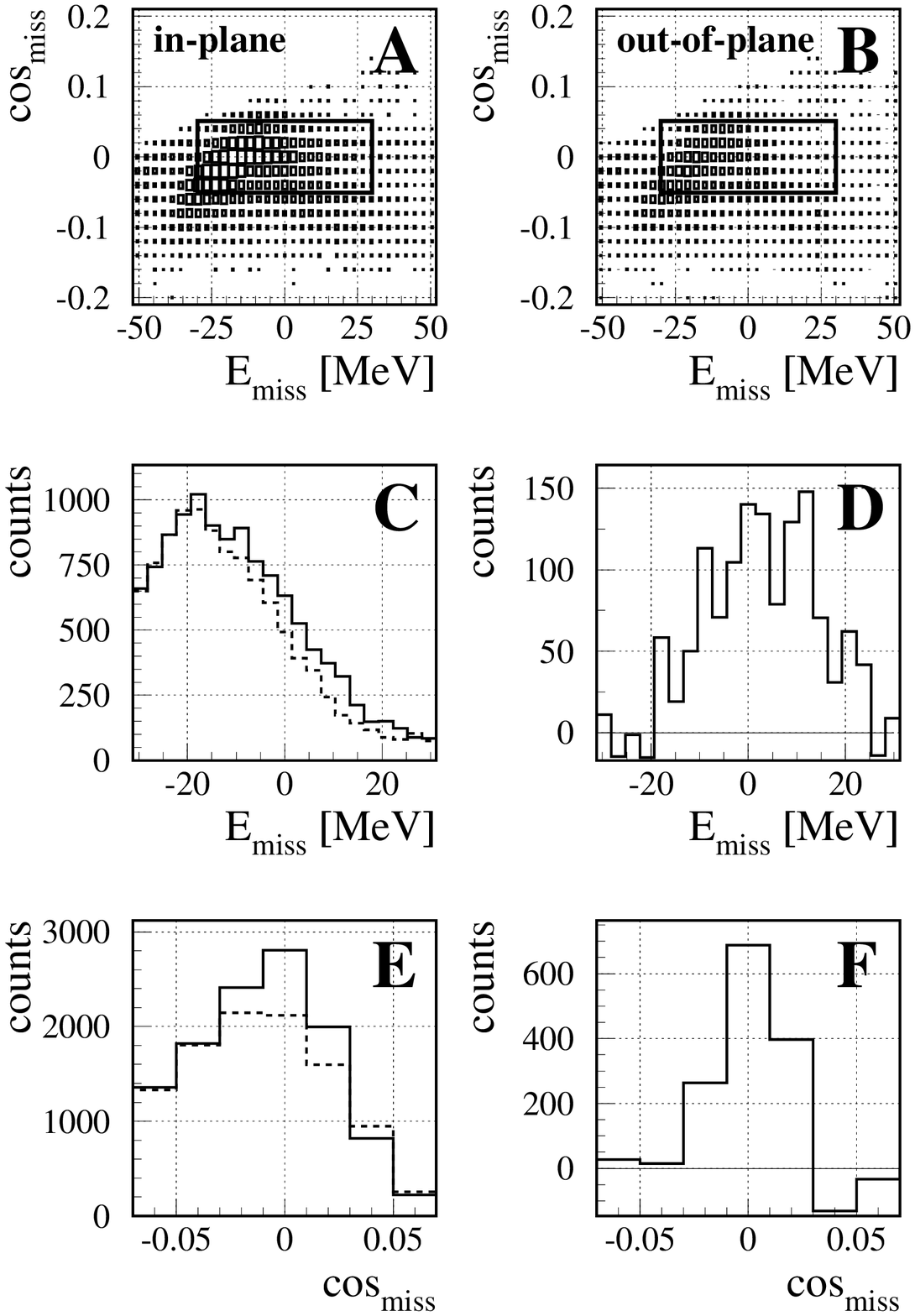}
  \caption{Two-dimensional analysis of experimental data obtained at a photon
           energy $E_\gamma = 779.9$ MeV and a scattering angle of
           $\theta^\mathrm{lab}_{\gamma}=32^\circ$. The data were obtained
           with 4 TOF plastic scintillator bars positioned at proton angles
           around $\theta^\mathrm{lab}_\mathrm{p} = 62^\circ$. For further
           details see Fig.~\ref{fig:fig6}.}
  \label{fig:fig8}
\end{figure}
Protons were identified through their comparatively large energy-deposition 
in a TD detector and through their time-of-flight.  
For each proton event detected
by a TOF detector a trajectory was constructed using the intersection points
in the two wire chambers. The event was accepted as a good one if the
trajectory intersected the scattering target, hit the appropriate TD detector
and intersected the TOF detector at the experimental impact point within its
spatial resolution.  Then, for a given proton trajectory and a given primary
photon energy $E_{\gamma}$ the direction $\theta^\mathrm{Comp}_{\gamma}$ and
energy $E^\mathrm{Comp}_{\gamma}$ of the secondary photon as well as the
energy $E^\mathrm{Comp}_\mathrm{p}$ of the recoil proton were calculated
assuming Compton kinematics.  Only those events were accepted where the
experimental direction of the secondary photon was close to the direction
calculated for a Compton photon. This procedure led to a drastic reduction of
the number of background events from $\pi^0$ photoproduction.

The final separation of events from Compton scattering and $\pi^0$
photoproduction was achieved by time-of-flight analysis.  The experimental
time-of-flight was compared with the one calculated from the energy
$E^\mathrm{Comp}_\mathrm{p}$ expected for a recoil proton of a Compton event.
Mean energy losses of the proton were used in this calculation. The difference
between the experimental and the calculated time-of-flight was named the
missing time $\Delta t_\mathrm{p}$.

Figs.~\ref{fig:fig4} and \ref{fig:fig5} show typical missing time spectra for
incident photon energies of $E_\gamma$ = 345.3 MeV, 413.0 MeV and 659.3 MeV,
the former two for intermediate photon angle of $\theta^\mathrm{lab}_\gamma =
70^\circ$ and the latter for a large photon angle of $\theta^{\rm lab}_\gamma
= 116^\circ$.  The corresponding proton angles were
$\theta^\mathrm{lab}_\mathrm{p} = 45^\circ$ and $20^\circ$, respectively.
These three cases were selected to demonstrate examples of ``comparatively
easy'' separation of events from the $(\gamma,\gamma)$ and $(\gamma,\pi^0)$
reactions.  At the lowest energy of $E_\gamma$ = 345.3 MeV there is a complete
separation of the two types of events, whereas at the higher energy of
$E_\gamma$ = 413.0 MeV there is some overlap which can be removed by
subtracting the tail of the $(\gamma,\pi^0)$ events underneath the
$(\gamma,\gamma)$ events. The shape of this tail was taken from the
out-of-plane data.  At the higher energy of $E_\gamma$ = 659.3 MeV the overlap
of the two types of events is complete. However, after using the appropriate
cuts the remaining background of $(\gamma,\pi^0)$ events is considerably
smaller than the corresponding number of $(\gamma,\gamma)$ events.  This made
the separation of the two types of events precise and comparatively easy. In
this case the background due to $(\gamma,\pi^0)$ events was taken from
experimental data where the photon was detected outside the Compton scattering
plane.  These out-of-plane data were then transferred into the Compton
scattering plane by help of the predictions of a computer simulation. 
This method
was already  successfully applied in one of the previous experiments
carried out in Mainz \cite{molinari96}. The validity of this method
was clearly demonstrated in Refs. \cite{molinari96} and \cite{huenger97}.

To determine the detector efficiencies the analysis of the experimental data
was accompanied by a Monte Carlo simulation taking into account all relevant
effects.  All calibrations needed as
inputs for a precise simulation, including the efficiencies of the wire
chambers, were found in a self-calibration procedure making use of the large
amount of data from the ($\gamma,\pi^0$) reaction.

In a second analysis of the data of  the second resonance region carried
out independently of the one described above,
the one-dimensional analysis in terms of the missing time $\Delta
t_\mathrm{p}$ was replaced by a two-dimensional 
analysis with
$\Delta t_\mathrm{p}$ -- or the equivalent missing energy E$_\mathrm{miss}$
-- and the difference $\cos_\mathrm{miss}$ between the experimental $\cos
\theta_{\gamma}$ and $\cos \theta^\mathrm{Comp}_{\gamma}$ as the two
coordinates.  The two dimensional procedure is illustrated in
Figs.~\ref{fig:fig6}--\ref{fig:fig8} corresponding to a small photon angle in
the range of $\theta^\mathrm{lab}_\gamma$ = 28$^\circ$--37$^\circ$ where the
separation of the two types of events is ``comparatively difficult''. The two
upper subfigures (A) and (B) show scatter plots of events, with
the photon detected in the Compton scattering plane and 
outside the Compton scattering plane, respectively.
The rectangular frames are chosen such that for the in-plane data (subfigures
(A)) Compton events are entirely located in this frame. By comparing the
Figs.~\ref{fig:fig6}--\ref{fig:fig8} with each other we notice that the peaks
of the $(\gamma,\pi^0)$ distributions are outside the rectangular frames at the
lowest photon energy and are moving  into the center of the rectangular frames
at the highest photon energy. This is in line with the expectation that with
increasing photon energy effects of the finite pion mass become less
important.  As before, the background from $\pi^0$ photoproduction was
obtained from the out-of-plane data and subtracted from the in-plane data.
For this procedure the scatter plots of $(\gamma,\pi^0)$ events in the two
upper parts of Figs.~\ref{fig:fig6}--\ref{fig:fig8} were also generated by a
Computer simulation and adjusted to the corresponding experimental data
outside the rectangular frames. These adjusted simulated data were then used
to correct for possible differences in the experimental $(\gamma,\pi^0)$ data
located in the rectangular frames of the subfigures (A) and (B).  In
subfigures (C) to (F) projections of the data located inside the rectangular
frames on the $E_\mathrm{miss}$ and $\cos_\mathrm{miss}$ axes, respectively,
are shown.  In the subfigures (C) and (E) the solid curves represent
$(\gamma,\gamma)$ plus $(\gamma,\pi^0)$ events (in-plane data) and the dashed
curves the $(\gamma,\pi^0)$ background (out-of-plane data). The curve in
the subfigures (D) and (F) show the net number of $(\gamma,\gamma)$ events.
The projections in the subfigures (C) to (F) of
Figs.~\ref{fig:fig6}--\ref{fig:fig8} are shown for illustration, whereas the
differential cross sections for Compton scattering have been derived by
directly evaluating the contents of the rectangular frames of the upper
subfigures (A) and (B).  This two-dimensional analysis extended the available
differential cross sections to smaller scattering angles as compared to the
one-dimensional analysis. The results nicely agree with those from the
one-dimensional analysis in the regions where both types of analyses have been
carried out.

The procedures described above led to data with individual (random)
errors which have been carefully determined during the evaluation 
procedure.  These random errors are due to the counting statistics 
and the systematic errors due to the detection efficiency, 
the geometrical uncertainty of the apparatus and of
the background-subtraction procedure. There are  additional common (sca\-le)
systematic errors due to the tagging efficiency $(\pm 2\%)$ and 
target density and thickness $(\pm 2\%)$. The scale
errors of the quantities extracted from our data were obtained by scaling all
data points to 97\% and 103\% of their nominal values. 
Since the random
errors contain statistical and systematic components we do not discriminate
between these two types of errors in the results presented in the following.
The combined statistical+systematic errors have been obtained by adding 
random and scale errors in quadrature.

The number of differential cross sections obtained for the first resonance
region below 455 MeV is 436. With the two different analyses a total number
of 329 differential cross sections has been obtained for the second resonance
region above 455 MeV.  Of these 221 are partly overlapping with respect to the
energy and angular range. This overlap has carefully been taken into account
in the determination of the number of degrees of freedom (d.o.f.) used in the
$\chi^2$ procedures described in the following. Since it appeared
inappropriate to combine two data points from only partly overlapping
intervals into one data point by averaging, the following procedure was
applied. The two data points were kept separate but their individual errors
were enlarged by a factor of $\sqrt{2}$, giving a hypothetical arithmetic
average the same error as the single data points have.

The differential cross sections obtained in the present experiment are given
in Tables \ref{tab:tab1} to \ref{tab:tab3} shown in the appendix.
%%%%%%%%%%%%%%%%%%%%%%%%%%%%%%%%%%%%%%%%%%%%%%%%%%%%%%%%%%%%%%%%%%%
%%%
%%%     S E C T I O N       Theory
%%%
\section{Theory}
\label{sec:theory}
In the general case Compton scattering is described by six invariant amplitudes
$A_i(\nu,t)$, $i=1\cdots 6$  \cite{lvov97}  where 
\begin{eqnarray}
\nu&=&\frac{s-u}{4m}=E_\gamma+\frac{t}{4m},\quad t=(k-k')^2,\quad s=(k+p)^2,
\nonumber\\
u&=&(k-p')^2
\label{T4}
\end{eqnarray}
and $s+u+t=2m^2$. These amplitudes can be constructed to have no kinematical
singularities and constraints and to obey the usual dispersion relations.
We formulate fixed-$t$ dispersions relations for $A_i(\nu,t)$ by using a
Cauchy loop of finite size (a closed semicircle of radius $\nu_{max}$),
so that
\begin{equation}
{\rm Re} A_i(\nu,t)=A^{\rm pole}_i(\nu,t) 
+A^{\rm int}_i(\nu,t)+A^{\rm as}_i(\nu,t)
\label{T5}
\end{equation}
with 
\begin{eqnarray}
A^{\rm pole}_i(\nu,t) &=&\frac{a_i(t)}{\nu^2-t^2/16m^2}\nonumber\\
A^{\rm int}(\nu,t)&=&\frac{2}{\pi}{\cal P}\int^{\nu_{\rm max}(t)}_{\nu_{\rm thr}
(t)} {\rm Im} A_i(\nu',t)\frac{\nu' d\nu'}{{\nu'}^2 - \nu^2}\nonumber\\
A^{\rm as}(\nu,t)&=&\frac{1}{\pi}{\rm Im}\int_{{\cal C}_{\nu_{\rm max}}}
A_i(\nu',t)\frac{\nu' d\nu'}{{\nu'}^2 -\nu^2}.
\label{T6}
\end{eqnarray}
The explicit use of the contour integral for $A^{\rm as}(\nu,t)$
is only necessary for i = 1 and 2, where special models have to be used 
for this purpose. For i = 3 $\cdots$ 6 the contour integral for 
$A^{\rm as}(\nu,t)$ can be avoided by extending the integral for
$ A^{\rm int}(\nu,t)$ to infinity. 
  
The integral contributions $A^{\rm int}_i(\nu,t)$ are 
determined by the imaginary
part of the Compton scattering amplitude which is given by the unitarity
relation of the generic form
\begin{equation}
2 {\rm Im} T_{fi}= \sum_n (2\pi)^4\delta^4(P_n-P_i)T^*_{nf}T_{ni}.
\label{T7}
\end{equation}   
The quantities entering into the r.h.s. of (\ref{T7}) are from $n=\pi N$
and $n=2\pi N$ intermediate states where the $n=\pi N$ component can be
constructed from  parameterizations of pion 
photoproduction  multipoles $E_{l\pm}$, $M_{l\pm}$. The  $n=2\pi N$
component requires additional model considerations \cite{lvov97}.

For the asymptotic part of the amplitude $A_2(\nu,t)$ we may use the Low
amplitude of the $\pi^0$ exchange in the $t$-channel
\begin{equation}
A^{\rm as}_2(t)\simeq A^{\pi^0}_2(t)=\frac{g_{\pi NN}F_{\pi^0 \gamma\gamma}}
{t-m^2_{\pi^0}}\tau _3 F_\pi (t),
\label{T8}
\end{equation}
where the isospin factor is $\tau_3=\pm 1$ for the proton and neutron,
respectively, and the product of the $\pi NN$ and $\pi^0 \gamma\gamma$
couplings is
\begin{eqnarray}
g_{\pi NN}F_{\pi^0 \gamma\gamma}&=&-16 \pi\sqrt{\frac{g^2_{\pi NN}}{4\pi}
\frac{\Gamma_{\pi^0\to 2\gamma}}{m^3_{\pi^0}}}\nonumber\\
&=&(-0.331\pm 0.012)\,{\rm GeV}^{-1}.
\label{T9}
\end{eqnarray}
The inclusion of small corrections due to the $\eta$ and $\eta'$ mesons
has been described elsewhere \cite{lvov99}.
There may be  arguments that $A^{\rm as}_2(t)$ is not exhausted by
$\pi^0$ exchange in the $t$-channel. In order to introduce an additional
parameter into the relevant amplitude which provides the necessary flexibility
for an experimental test we write 
\begin{equation}
A^{\rm as}_2(t)\simeq A^{\pi^0}_2(t)- 2\pi m\frac{\delta\gamma_\pi}{
1-\frac{t}{\Lambda^2}}
\label{T9a}
\end{equation}
from which the substitution follows:
\begin{equation}
\gamma_\pi \to \gamma_\pi +  \delta\gamma_\pi.
 \label{T9b}
\end{equation}
The parameter $\Lambda$ defines the slope of the function at $t = 0$ and 
is chosen to be $\Lambda$ = 700 MeV. In varying $\delta\gamma_\pi$ the
influence of any deviation from the standard value of $\gamma_\pi$
can be investigated in terms of this ansatz.

The asymptotic contribution of the amplitude $A_1(\nu,t)$ is modeled
through an ansatz analogous  to the Low amplitude, except for the fact
that the pseudoscalar meson $\pi^0$ is replaced by the  scalar $\sigma$ 
meson. In this case we use a simpler form  of the ansatz
\begin{equation}
A^{\rm as}_1(t)\simeq A^\sigma_1(t)=\frac{g_{\sigma NN}F_{\sigma\gamma\gamma}}
{t-m^2_\sigma}
\label{T10}
\end{equation}
and include quantities like the formfactor in (\ref{T8}) into the 
"effective mass" $m_\sigma$ being  now  an adjustable 
parameter \cite{lvov97,galler01}. 
The quantity
$g_{\sigma NN}F_{\sigma \gamma\gamma}$ is given by the difference of 
the electric and magnetic polarizabilities $\alpha-\beta$ through
\begin{equation}
2 \pi(\alpha-\beta)+A^{\rm int}_1(0,0)=-A^{\rm as}_1(0,0)=
\frac{g_{\sigma NN} F_{\sigma\gamma\gamma}}{m^2_\sigma}
\end{equation} 
with the integral part being a minor contribution.
Though this $\sigma$-pole ansatz proved to be very 
successful \cite{lvov97,galler01} when compared with experimental data,
it would be desirable to have an independent  justification 
through an investigation of the
relevant $t$-channel. Studies of this type are in progress.

%%%%%%%%%%%%%%%%%%%%%%%%%%%%%%%%%%%%%%%%%%%%%%%%%%%%%%%%%%%%%%%%%%%
%%%
%%%     S E C T I O N       Results and Discussion
%%%
\section{Results and Discussion}
\label{sec:resultsanddiscussion}
\begin{figure*}[t]\sidecaption
  \centering \includegraphics[width=0.65\textwidth]{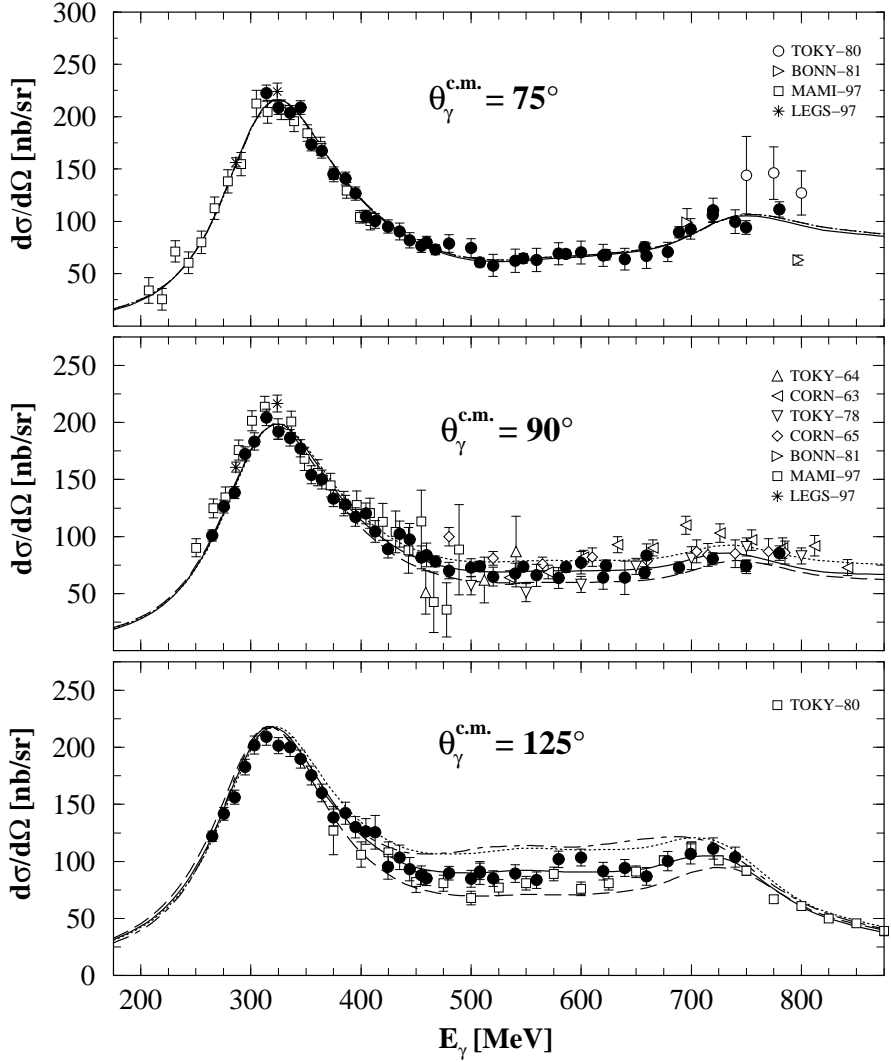}
  \caption{Differential cross sections for the first and second resonance
           region in comparison with data from other experiments. The curves
           show calculations based on the SAID-SM99K photo-meson amplitudes
           for $m_\sigma = 400 \; \mathrm{MeV}$ (dashed),
           $m_\sigma = 600 \; \mathrm{MeV}$ (solid) and
           $m_\sigma = 800 \; \mathrm{MeV}$ (dotted). Other parameters
           are those in Eq. (\ref{freevariation}). The dashed-dotted
           curve given for the angle 125$^\circ$ shows calculations based on
           the MAID2K photo-meson amplitudes with $m_\sigma=$ 600 MeV
           and the other parameters specified in Eq. (\ref{fixed}).
           The previous data are
           compiled in \cite{ukai97} and are taken from:
           \protect\cite{ishii80} (TOKY-80);
           \protect\cite{jung81} (BONN-81);
           \protect\cite{huenger97} (MAMI-97);
           \protect\cite{tonnison98} (LEGS-97);
           \protect\cite{nagashima64} (TOKY-64);
           \protect\cite{stiening63} (CORN-63);
           \protect\cite{toshioka78} (TOKY-78);
           \protect\cite{rust65} (CORN-65). The data of the present work
           (filled circles, representing angular intervals of 
           $\Delta \theta^{c.m.}_\gamma = 15^\circ$) are given
           with error bars
           taking into account the counting statistics, and systematic errors
           due to detection efficiency, geometrical uncertainties and
           background subtraction.}
  \label{fig:fig9}
\end{figure*}
\begin{figure*}[t]\sidecaption
  \centering \includegraphics[width=0.65\textwidth]{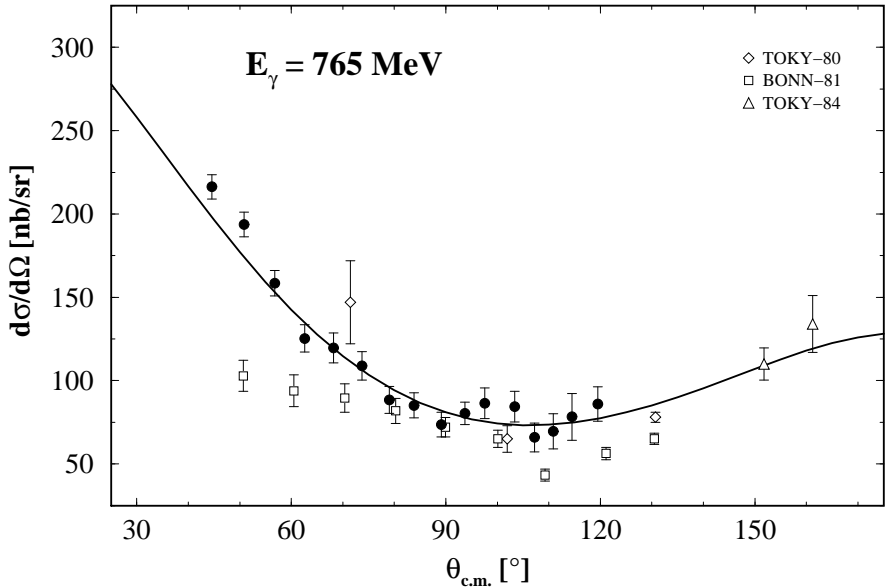}
  \caption{Differential cross sections for 765 MeV  photon energy
           in comparison with data from other experiments. The curve shows a
           calculation based on the SAID-SM99K parameterization and on the
           parameters given in (\ref{freevariation}).
           The previous data are taken from:
           \protect\cite{jung81} (BONN-81);
           \protect\cite{ishii80} (TOKY-80);
           \protect\cite{wada84} (TOKY-84). The
           data from the present work ({\tiny \ding{108}}) represent energy
           intervals of $\Delta E_\gamma$ = 60 MeV width. Their 
           error bars take into
           account the counting statistics and systematic errors due to
           detection efficiency, geometrical uncertainties and background
           subtraction.}
  \label{fig:fig10}
\end{figure*}
\begin{figure*}[t]\sidecaption
  \centering \includegraphics[width=0.65\textwidth]{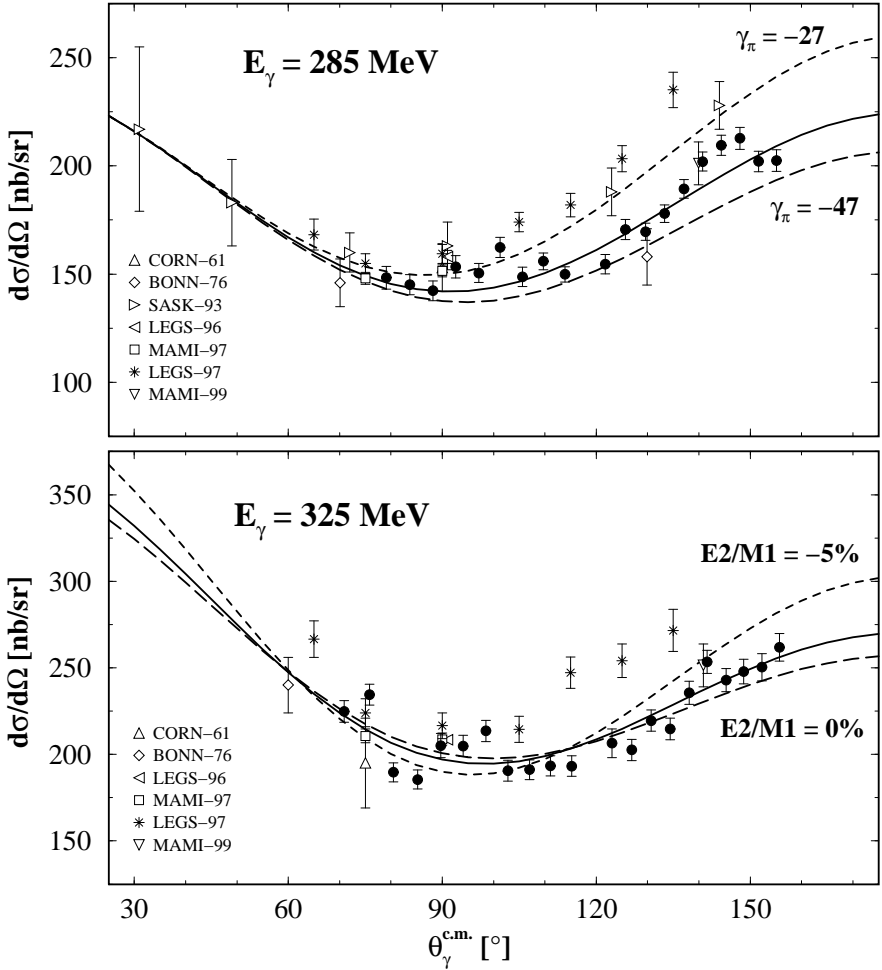}%% fig11
  \caption{Angular distributions of Compton differential cross sections
           obtained with the LARA arrangement (filled circles, representing
           energy intervals of $\Delta E_\gamma =$ 40 MeV) 
           compared with previous data
           as compiled in \protect\cite{ukai97} and with predictions of 
           dispersion
           theory with the SAID-SM99K photo-pion amplitudes. The full lines
           are obtained if the parameters of 
           Eq. (\ref{freevariation}) are applied. The dashed lines show 
           sensitivities to $\gamma_\pi$ at $E_\gamma =$ 285 MeV (upper part)
           and to the ratio E2/M1 at $E_\gamma =$ 325 MeV (lower part).  
           The previous data are from: \protect\cite{dewire61} (CORN-61);
           \cite{hallin93} (SASK-93); \protect\cite{genzel76} (BONN-76);
           \cite{blanpied96} (LEGS-96); \protect\cite{huenger97} (MAMI-97);
           \cite{tonnison98} (LEGS-97); \protect\cite{wissmann99}
           (MAMI-99). The final  value for the parameter 
           $\gamma_\pi$ has not been obtained from these data points only
           but from the total amount of
           data below 455 MeV (see the text).}
  \label{fig:fig11}
\end{figure*}
\begin{figure*}[t]\sidecaption
  \centering \includegraphics[width=0.65\textwidth]{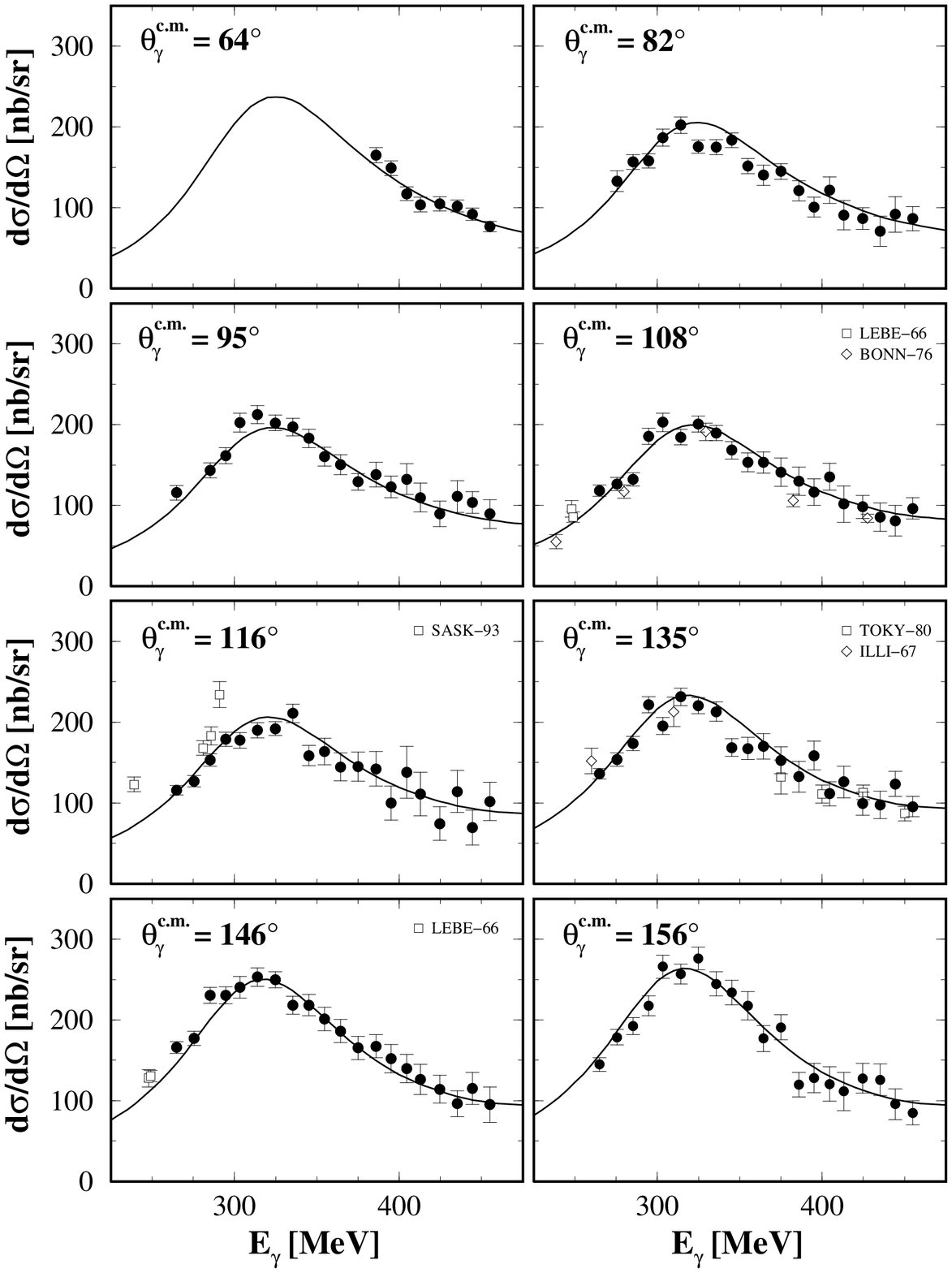}%%fig12
  \caption{Eight out of 24 energy distributions from $59^\circ$
           to $156^\circ$ (c.m.) obtained with the LARA arrangement in the
           first resonance range compared with previous data and with
           predictions from dispersion theory. The previous data are taken
           from:
           \protect\cite{baranov66a,baranov66b} (LEBE-66);
           \protect\cite{gray67} (ILLI-67);
           \protect\cite{genzel76} (BONN-76);
           \protect\cite{hallin93} (SASK-93);
           \protect\cite{ishii80} (TOKY-80);
           the present work ($\bullet$) with error bars as in
           Fig.~\ref{fig:fig9}.}
  \label{fig:fig12}
\end{figure*}
In Figs.~\ref{fig:fig9}--\ref{fig:fig12} we discuss specific properties
of our present experimental data in comparison with predictions and 
with previous results. 
Figs.~\ref{fig:fig13}--\ref{fig:fig17} in the appendix show the complete
set of data obtained in the present experiment compared with the same kind of
predictions. 
As predictions we use the results of the dispersion theory \cite{lvov97} 
based on the SAID-SM99K
parametrization of photo-meson amplitudes \cite{arndt96} together with the
parameter $\alpha - \beta$, the difference of the electric and magnetic 
polarizability. For the latter quantity the global average 
$\alpha - \beta = (10.0 \pm 1.5_\mathrm{stat+syst} \pm
0.9_\mathrm{model}) \times 10^{-4} \; \mathrm{fm}^3$ has been obtained,
taking into account  experiments
of 90's \cite{macgibbon95}.\footnote{The use of a twice as large data base of
  50's--90's and a fit without the Baldin sum rule constraint leads to $\alpha
  = (11.7 \pm 0.8_\mathrm{stat+syst} \pm 0.7_\mathrm{model}) \times$ $ 10^{-4}
  \; \mathrm{fm}^3$ and $\beta = (2.3 \pm 0.9_\mathrm{stat+syst} \pm
  0.7_\mathrm{model}) \times$ $ 10^{-4} \; \mathrm{fm}^3$ \cite{baranov00} and
  thus confirms the above finding.}  More recently the LEGS group
\cite{tonnison98}
published the result $\alpha-\beta =10.11\pm 1.74_{\rm stat+syst}$
and the TAPS collaboration at MAMI (Mainz) obtained a new global average
of $\alpha -\beta=10.5\pm 0.9_{\rm stat+syst}\pm 0.7_{\rm model}$ 
\cite{olmos01}.

The parameter $\alpha - \beta$ was not
adjusted to the present data for two reasons: (i) This quantity is mainly due
to a $t$-channel exchange and, therefore, essentially independent of the
parametrization of photo-meson amplitudes. (ii) This quantity is strongly
constrained by large-angle differential cross sections below pion
photoproduction threshold where the present experiment made no contribution.
The total photoabsorption cross section corresponding to the presently used
parametrization leads through the Baldin sum rule to $\alpha + \beta = 14.05
\times 10^{-4} \; \mathrm{fm}^3$ and is in between the values of Babusci et
al.\ \cite{babusci98}, being  $\alpha + \beta = \left( 13.69 \pm 0.14
\right) \times 10^{-4} \; \mathrm{fm}^3$, and $\alpha + \beta = \left( 14.2
  \pm 0.3 \right) \times 10^{-4} \; \mathrm{fm}^3$, which is based on
numerical results of Ref.~\cite{damashek70}. Some critical discussion of these
and related numbers can be found in \cite{levchuk00}.

From the present data in the second resonance region the only remaining free
parameter of the dispersion theory \cite{lvov97}, the effective mass-parameter
of the $\sigma$ meson,  was fitted leading to $m_\sigma = (589 \pm
12_\mathrm{stat + syst})$ MeV with $\chi^2/{\rm d.o.f.} = 1.33$
which essentially confirms the previous estimate \cite{lvov97}
of $m_\sigma = 600$ MeV. 
The procedure is illustrated in Fig.\,\ref{fig:fig9} where the 
three curves have been calculated with the
effective mass parameters  $m_\sigma$ =400, 600 and 800 MeV. 
This Figure as well as the corresponding
data shown in Figs.~\ref{fig:fig15}-\ref{fig:fig17} of the  Appendix 
prove that the parametrization of the asymptotic
part of the invariant aplitude $A_1(\nu,t)$ introduced in \cite{lvov97}
 is in line with the experimental data.
The  present and previous \cite{lvov97} result of $m_\sigma = 600$ MeV
is in agreement with what is frequently
denoted as the ``mass of a sigma meson''. However, we wish to stress here that
we do not claim to have determined a ``mass of a sigma meson''. For us this
quantity merely is a number in the pole parametrization of the $t$-channel
$J^{PC}=0^{++}$ exchange in the region of negative $t$ which leads to an
excellent representation of the data of the second resonance region
\cite{lvov97}.
The data from the present and previous experiments shown in 
Fig.~\ref{fig:fig9} are in a general good  agreement with each other.
Nevertheless, the improvement in accuracy achieved in the present experiment
is quite apparent.

Systematic differences between present and previous data are seen in
in Fig.~\ref{fig:fig10} where the angular distribution of
differential cross sections is  shown for the  photon energy
E$_\gamma=765$ MeV. Here the data from the Bonn-81
experiment \cite{jung81} are considerably below our data and below the 
predictions,  especially in the forward direction. This shows
that the coverage of the second resonance region through data from previous
experiments was by far not sufficient.

After fixing  the effective mass-parameter 
$m_\sigma$  to 600 MeV  it is possible to use the differential
cross-sections in the first resonance region up to 455 MeV photon energy to
get information on two important quantities which were subject to several
recent investigations. These are the backward spin polarizability $\gamma_\pi$
and the E2/M1 ratio of the $p \to \Delta$ transition. In accordance with
previous work \cite{beck97a,beck00} the $\mathrm{E2/M1}$ ratio is 
defined here as the
ratio $\mbox{Im} E^{(3/2)}_{1+}/\mbox{Im}M^{(3/2)}_{1+}$ taken at the resonance
point\footnote{A small shift of this
energy leads to a significantly different value of the
E2/M1-ratio \cite{beck00}. This is of importance since the SAID-SM99K
parametrization favors a resonance point of $E_\gamma \approx 337$ MeV.} 
$E_\gamma= 340$ MeV \cite{beck97a,beck00}, 
where $\delta_{33}=90^\circ$
or, equivalently, $\mbox{Re}M^{(3/2)}_{1+}=0$. One can make a small change in
the $\Delta$-resonance contribution to $M^{(3/2)}_{1+}$ or $E^{(3/2)}_{1+}$ and
thus change the $\mathrm{E2/M1}$ ratio using a fine tuning of the
$\Delta$-resonance photocouplings $(M_{1+}^{(3/2)})_r$ and $(E_{1+}^{(3/2)})_r$
as described elsewhere \cite{huenger97}. Such changes affect the imaginary part
of the Compton scattering amplitude \cite{huenger97} and, through the
dispersion relations, the real part too.
A similar procedure may be applied
to the backward spin polarizability $\gamma_\pi$ by adding an extra term to the
asymptotic contribution $A_2^{\rm as}$ (\ref{T9a}) usually represented 
only by the
$\pi^0$-exchange \cite{tonnison98}. Such a change affects the real part of the
Compton scattering amplitude only. 

For a given $M_{1+}$ amplitude which essentially fixes the predicted
differential cross sections at 
$\theta^\mathrm{c.m.}_\gamma=60^\circ$ and 
$115^\circ$, the E2/M1 ratio 
shows its highest sensitivity to the differential
cross sections in the maximum of the $\Delta$-resonance and for $90^\circ$ and
forward and backward angles.  In practice this  procedure gains its highest
sensitivity if it is restricted to the subset of data between 280 and 360 MeV. 
The backward spin polarizability
$\gamma_\pi$ shows its highest sensitivity to the differential cross sections
for beam energies of about 285 MeV and only in the backward direction.
In this case the evaluation may be caried out using all data below 455 MeV.  
The sensitivity of the data to the quantities $\gamma_\pi$ and E2/M1 
is illustrated in the lower and upper parts of Fig.\,~\ref{fig:fig11},
respectively. 

The overall quality of the data 
obtained in the present experiment for the first resonance region may be 
judged from Fig.~\ref{fig:fig12} which  shows  
selected examples of differential cross sections. There is a general 
good agreement with previous data with only few exceptions. 
This figure shows that in
the $\Delta$ energy range the coverage with experimental differential cross
sections is good except for small angles. The reason for this lack of data at
small angles is that the recoil proton has a too low energy to leave the
scattering target.  In this range additional data may be measured using the
large Mainz NaI(Tl) detector without recoil-proton detection.

In detail we used the following procedure to determine the multipoles  
characterizing  the $\Delta$-resonance and to
extract $\gamma_\pi$:  We start with the
fixed  mass parameter $m_\sigma=600$ MeV and the new global
average for the difference of the
electric and magnetic polarizabilities of the proton \cite{olmos01},
$\alpha-\beta=(10.5\pm 0.9_{\rm stat+syst}\pm 0.7_{\rm model})
\times 10^{-4}~\rm fm^3$ which nicely confirms the previous 
one \cite{macgibbon95} but with a reduced experimental error.
Taking a
subset of 167 data points close to the $\Delta$-resonance peak, namely those
between the limits $E_\gamma = 280$ and 360 MeV where the
$\Delta$-resonance contribution strongly dominates, we slightly rescale the
$\Delta$-resonance parts of the photo-pion amplitudes $M_{1+}$ and
$E_{1+}$, as described in \cite{huenger97}, in order to achieve the
best agreement between the present experimental data and dispersion-theory
predictions. The above
choice of the energy limits is made in order to reduce otherwise bigger
model errors in the determination of the resonance parameters.
With these corrected amplitudes, setting an overall scale for the
theoretical differential cross sections of Compton scattering close to
the resonance, we tune $\gamma_\pi$ through the asymptotic contribution
to the invariant amplitude $A_2$   (\ref{T9a})   in order to
arrive at the best $\chi^2$ in the whole energy region covering the
$\Delta$-resonance, which here is the region $E_\gamma \le 455$ MeV
containing 467 data points.
With this $\gamma_\pi$ we repeat the determination of the amplitudes
$M_{1+}$ and $E_{1+}$
and then arrive  again at $\gamma_\pi$, etc. These iterations quickly converge 
and eventually give the final values  for $M_{1+}$, $E_{1+}$ and
$\gamma_\pi$.

In order to determine the  model uncertainties of  the extracted quantities
we used different values for $\alpha-\beta$ within the experimental 
uncertainty of this quantity (i.e.\
between 9.4 and $11.6\times 10^{-4}~\rm fm^3$
\cite{olmos01}).
Also different values for $m_\sigma$ were used between 500 to 700 MeV. 
This range of $m_\sigma$ is supported by a comparison of different theoretical
calculations of the amplitude $A_1$ 
\cite{lvov97,drechsel99,holstein94,lvov99a}.
Moreover, we varied the $\pi^0\gamma\gamma$ coupling by $\pm$4\% 
and the $\eta NN$ and $\eta'NN$ couplings by $\pm$50\%.
The form factors accompanying the $\pi^0$, $\eta$, $\eta'$ $t$-channel
contributions were varied and also the parameters
which determine the multipole structure
of double-pion photoproduction below 800 MeV where the latter variation
was based on experience of a recent GDH experiment \cite{arends00}.

We present our findings in terms of the absolute value of the
$M_{1+}^{(3/2)}$ amplitude at the energy 320.0 MeV corresponding to the
maximum of  the differential cross section for  Compton scattering.
The E2/M1 ratio (EMR) of the imaginary parts of the amplitudes
$E_{1+}^{(3/2)}$ and $M_{1+}^{(3/2)}$  is determined
for  340.0 MeV where the real parts of these amplitudes are about
zero, in complete agreement with the previous procedure  
\cite{blanpied97,beck97a,beck00} where 
 the ratio of the imaginary parts was determined from pion
photoproduction experiments. It is important to exactly 
use the same energy $E_\gamma$ when comparing
the amplitudes $E_{1+}^{(3/2)}$ and $M_{1+}^{(3/2)}$ obtained 
from different experiments because they rapidly vary with  $E_\gamma$.
Our results are
\begin{eqnarray}
    |M_{1+}^{(3/2)}(320~{\rm MeV})|
      &=& (39.7 \pm 0.3_{\rm stat+syst} \pm 0.03_{\rm model})\nonumber\\
       & \times& 10^{-3} / m_{\pi^+} ,\nonumber\\
     {\rm EMR}(340~{\rm MeV})
      &=& (-1.7 \pm 0.4_{\rm stat+syst} \pm 0.2_{\rm model})\;\% ,
\nonumber \\
     \gamma_\pi
      &=& (-37.1 \pm 0.6_{\rm stat+syst} \pm 3.0_{\rm model})\nonumber\\
       & \times& 10^{-4} ~\rm fm^4.
\label{freevariation}
\end{eqnarray}
The systematic errors given here include changes imposed by 
a simultaneous shift
of all data points within the scale uncertainty of $\pm$3\%. This
uncertainty fully dominates the resulting uncertainty of  the
$M_{1+}^{(3/2)}$ amplitude.
Note that the required modifications of the  amplitudes $M_{1+}^{(3/2)}$
and $E_{1+}^{(3/2)}$ are 
compatible with zero. Without the modification,
the SAID-SM99K parameterization gives\\
$|M_{1+}^{(3/2)}(320~{\rm MeV})| = 39.74$ (in the same units) and\\
EMR$(340~{\rm MeV}) = -1.68\%$.
The present value for  $M_{1+}^{(3/2)}$ perfectly agrees with the one
previously determined by H\"unger et al.
\cite{huenger97}:  $|M_{1+}^{(3/2)}(320~{\rm MeV})| = 39.6 \pm 0.4$.
 Since 
we did not try
to tune other photo-meson amplitudes like $E_{0+}$, $M_{1-}$ or $E_{2-}$ which
are also of importance for a good description of Compton scattering data near
the $\Delta$-resonance, the model errors in (\ref{freevariation}) 
may still be incomplete.

The value of EMR determined from the present Compton scattering data
is  smaller than the one  obtained in a dedicated Mainz  photo-pion
experiment, i.e. $(-2.5\pm 0.1_{\rm stat} \pm 0.2_{\rm syst})\%$ 
\cite{beck97a,beck00},
and also smaller than the result published by the LEGS group
\cite{blanpied97}, i.e.
$(-3.0\pm 0.3_{\rm stat+syst}\pm 0.2_{\rm model})\%$.
Our result essentially confirms the prediction of the SAID-SM99K
parameterization, in agreement with the observation that this parameterization
leads to an overall agreement with our Compton scattering differential
cross sections. However, it should be noted that by applying 
the same procedure as before but fixing the E2/M1 ratio
to EMR(340 MeV)=$-2.5\%$, a good fit to our data in the $\Delta$ resonance
region may also be obtained with only  slight shifts in the parameters
$M_{1+}^{(3/2)}(320~{\rm MeV})$ and $\gamma_\pi$. Therefore, at this 
stage of the investigation we do not contribute to the extensive discussion
of the E2/M1 ratio \cite{beck97b,beck97c,davidson97,workman97} carried out in
the past.

The uncertainties of the spin polarizability $\gamma_\pi$
are dominated by the model errors, especially -- for a given choice
of photo-meson amplitudes -- by the variations of
$m_\sigma$ and $\alpha-\beta$. Taking these into account our 
result for $\gamma_\pi$ is in
disagreement with the one determined in 1997
by the LEGS group \cite{tonnison98} which gave the  smaller value
$\gamma_\pi^{\rm LEGS} = -27.1 \pm 2.2_{\rm stat+syst}{
^{+2.8}_{-2.4}}{}_{\rm model}$ (in the same
units of $10^{-4}~\rm fm^4$). This difference can be traced back to a
difference in the measured differential cross sections, as can be seen 
in Fig.~\ref{fig:fig11}. The former result  \cite{tonnison98}
is also  in contradiction to standard dispersion theory
\cite{drechsel98,babuscu98a,lvov99} and also to chiral perturbation theory
\cite{hemmert98,gellas00,kumar00}. As a consequence is was concluded that
hitherto unknown effects related to the spin structure of the nucleon might
exist. With our new data such effects are clearly ruled out, in accordance
with our recently published data on quasi-free scattering from the proton
\cite{wissmann99} and with the one
obtained very recently by the TAPS collaboration at MAMI \cite{olmos01}, 
i.e.
$\gamma_\pi=-36.1 \pm 2.1_{\rm stat} \mp 0.4_{\rm syst} \pm 0.8_{\rm model}$.

The present value of $\gamma_\pi \approx -37.1$ agrees well with
predictions of the unsubtracted dispersion relation for the invariant
amplitude $A_2$ adopted in \cite{lvov97}. The latter gives $-38.24$
with the same photo-meson input and with the same
energy cut in the dispersion integrals of  $E_{\rm max}=1.5$ GeV,
thus assuming no essential asymptotic
contributions beyond pseudoscalar-meson exchanges ($\pi^0,\eta,\eta'$).
The present value for  $\gamma_\pi$ satisfactorily agrees with predictions
of the ``small scale expansion" scheme, which effectively is
chiral perturbation theory including the $\Delta$-resonance,
$\gamma^{\rm SSE}_\pi = -37$ \cite{hemmert98}. It also agrees with standard
chiral perturbation theory to order $O(p^4)$, which does not include the
$\Delta$-resonance, $\gamma^{\rm ChPT}_\pi = -39$ \cite{kumar00},
provided $-45$ is used for the anomaly contribution to $\gamma_\pi$
from $\pi^0$ exchange\footnote
{We do not use another ChPT prediction,
$\gamma^{\rm ChPT}_\pi = -42$ \cite{gellas00} for reasons
explained in \cite{birse00}.}. Furthermore, it agrees
with backward-angle dispersion relations, which include the
$\Delta$ and the $\eta$-$\eta'$ exchanges,
$\gamma^{\rm DR}_\pi = -39.5 \pm 2.4$ \cite{lvov99}.
Thus, there is good overall consistency between the present
Compton scattering data, the dispersion theory,
and the SAID-SM99K photo-meson amplitudes.

Such a consistency is deteriorated when the latest SAID-SM00K
photo-pion amplitudes are used. This is because in that latest
parameterization the $M1$-strength of the $\Delta$-resonance
is decreased to $|M_{1+}^{(3/2)}(320~{\rm MeV})| = 39.16$.
Therefore, we have to increase the SM00K $M_{1+}{(3/2)}$- amplitude 
by $+1.2\%$
in order to achieve  a satisfactory description of Compton scattering.
When such a rearrangement is made, the value extracted for $\gamma_\pi$ 
is $\gamma_\pi = - 37.0$, i.e. it turns out to be only slightly
smaller than the one of  Eq.(1) with similar errors.

When using the  MAID2K \cite{maid} parameterization of
photo-pion amplitudes
the same procedure gives the results
\begin{eqnarray}
    |M_{1+}^{(3/2)}(320~{\rm MeV})|
      &=& (39.8 \pm 0.3_{\rm stat+syst} \pm 0.03_{\rm model})\nonumber\\
       & \times& 10^{-3} / m_{\pi^+} ,
\nonumber \\
     {\rm EMR}(340~{\rm MeV})
      &=& (-2.0 \pm 0.4_{\rm stat+syst} \pm 0.2_{\rm model})\;\% ,
\nonumber \\
     \gamma_\pi
      &=& (-40.9 \pm 0.4_{\rm stat+syst} \pm 2.2_{\rm model})\nonumber\\
       & \times& 10^{-4} ~\rm fm^4
      \label{fixed}
\end{eqnarray}
which are more at variance with Eq.~(\ref{freevariation}) 
than the alternatives discussed above.
In this case a slightly bigger rearrangement of the resonance
amplitudes is required in comparison with their original values
which, for MAID2K, are $|M_{1+}^{(3/2)}(320~{\rm MeV})| = 39.92$
and EMR$(340~{\rm MeV}) = -2.19\%$.
The biggest change is, however, in the spin polarizability
$\gamma_\pi$ which can be traced back to rather different
nonresonant amplitudes $E_{0+}$ and $E_{2-}$ in the SAID and MAID
representations in the $\Delta$-resonance range.
The overall quality of the description of the present Compton scattering data
at energies below 455 MeV,
containing 467 data points in total,
is approximately the same for the SAID and MAID photo-meson input.
The fitting procedure based on the two sets of parameterizations leads to
$\chi^2=564$ in both cases  and the differences in the
predictions are small  as can be seen in  Fig.~\ref{fig:fig9}. 

However, the properties of the SAID and MAID parameterizations are quite
different in the second resonance region.  
For instance, $\chi^2/{\rm d.o.f.}= 1.36$ is obtained for all data point above
455 MeV for the SAID-based theoretical predictions with SAID-based
parameters (1), whereas $\chi^2/{\rm d.o.f.} =2.00$ is obtained for 
the same data points
with MAID-based theoretical predictions and MAID-based parameters (2).
This means that the MAID-based parameterization does not lead to a reasonable
fit to the data when the same parameter $m_\sigma=600$ MeV is used.
The biggest difference between these two versions is seen at backward
angles in the dip region between the first and second nucleon resonance,
as illustrated by the dashed-dotted curve in Fig.~\ref{fig:fig9}.
The use of a smaller $m_\sigma$ with the {\em same} $\gamma_\pi$ 
reduces  the  discrepancy in the dip region, however without leading to an
overall agreement. It is observed that
the fit to the data  below 455 MeV carried out with that smaller
$m_\sigma$ requires an even bigger $-\gamma_\pi$ compared to the one given in
(2), and with this bigger
$-\gamma_\pi$ again no agreement is achieved between the theory and the data
in the dip region.

\section{Conclusions}
The results of the present experiment may be summarized as follows.
For the first time Compton scattering by the proton has been measured
with a large acceptance set-up for the scattering angle and the photon energy.
The data confirm the magnitude of the $M1$-strength adopted
in the SAID-SM99K and MAID2K parameterizations (not in SAID-SM00K),
and are in agreement with the E2/M1 ratio given by these
parameterizations.
The backward spin polarizability $\gamma_\pi$ is found
to be in agreement with latest theoretical calculations,
although model errors should yet be better understood.

%%%%%%%%%%%%%%%%%%%%%%%%%%%%%%%%%%%%%%%%%%%%%%%%%%%%%%%%%%%%%%%%%%%
%%%
%%%     S E C T I O N       Acknowledgment
%%%
\section*{Acknowledgement}
  The authors are greatly indebted to Professor Turleiv Buran, Department of
  Physics, University of Oslo and to the Norwegian Research Council for
  Science and the Humanities for having given us the opportunity to use their
  equipment of 120 lead glass detectors for this experiment containing
  150 of these lead glass detectors in total.

%%%%%%%%%%%%%%%%%%%%%%%%%%%%%%%%%%%%%%%%%%%%%%%%%%%%%%%%%%%%%%%%%%%
%%%
%%%     A N H A N G
%%%
%\appendix
%\section{Appendix}
%A compilation of the results taken with the present experiment is given
%in tables \ref{tab:tab1}--\ref{tab:tab16} and figures
%\ref{fig:fig13}--\ref{fig:fig17}.

%%%%%%%%%%%%%%%%%%%%%%%%%%%%%%%%%%%%%%%%%%%%%%%%%%%%%%%%%%%%%%%%%%%
%%%
%%%     T A B E L L E N    U N D    P L O T S
%%%
\begin{figure*}
  \centering \includegraphics[width=0.8\textwidth]{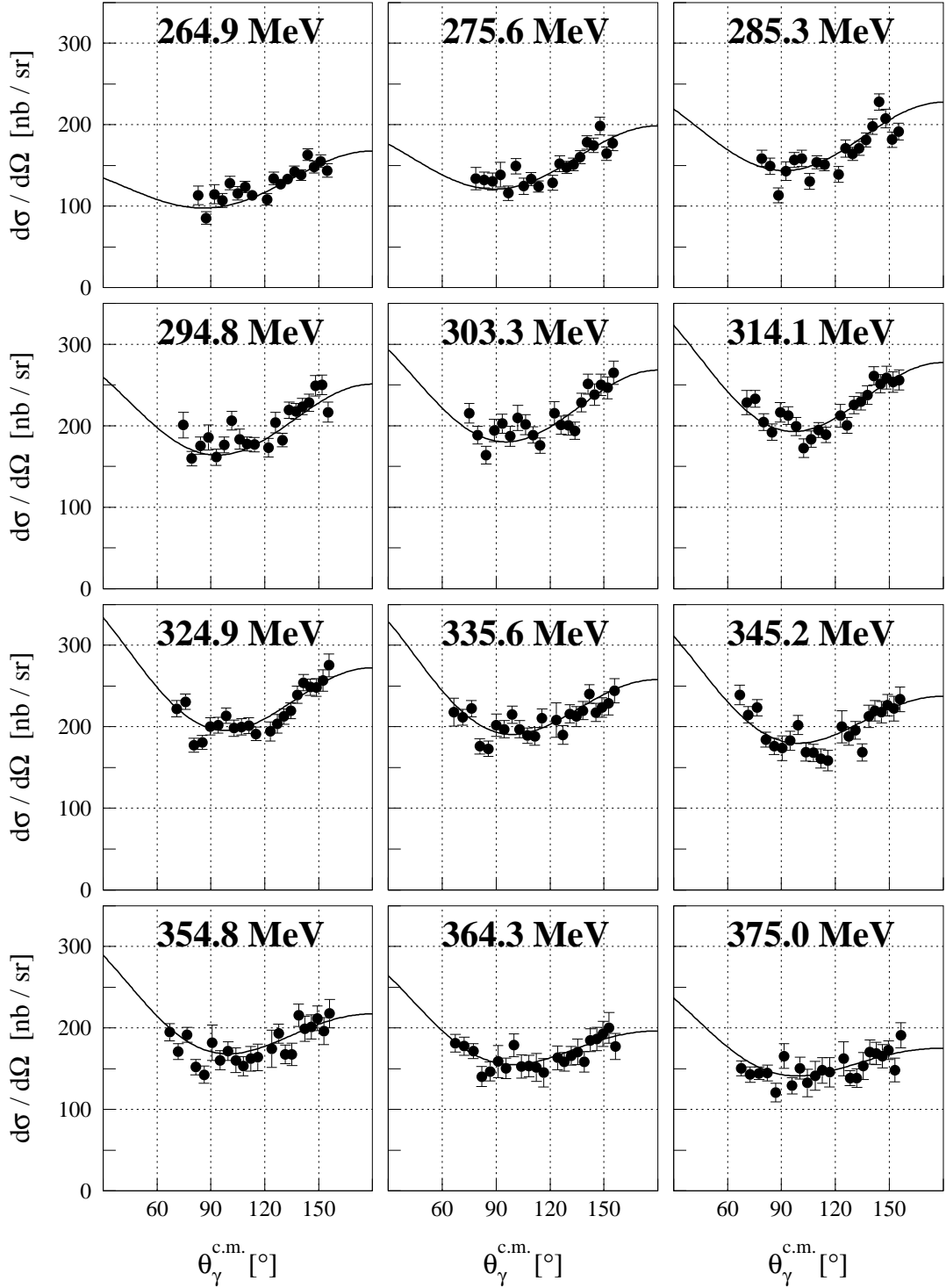}
  \caption{Angular distributions of the differential cross section in the
           c.m.-system as obtained with the LARA arrangement using the
           ``one-dimensional'' analysis. Solid line: calculation within the
           dispersion relation approach \protect\cite{lvov97} 
           using the 'best-fit'
           parameters of Eqs.~(\ref{freevariation}).}
  \label{fig:fig13}
\end{figure*}
\begin{figure*}
  \centering \includegraphics[width=0.8\textwidth]{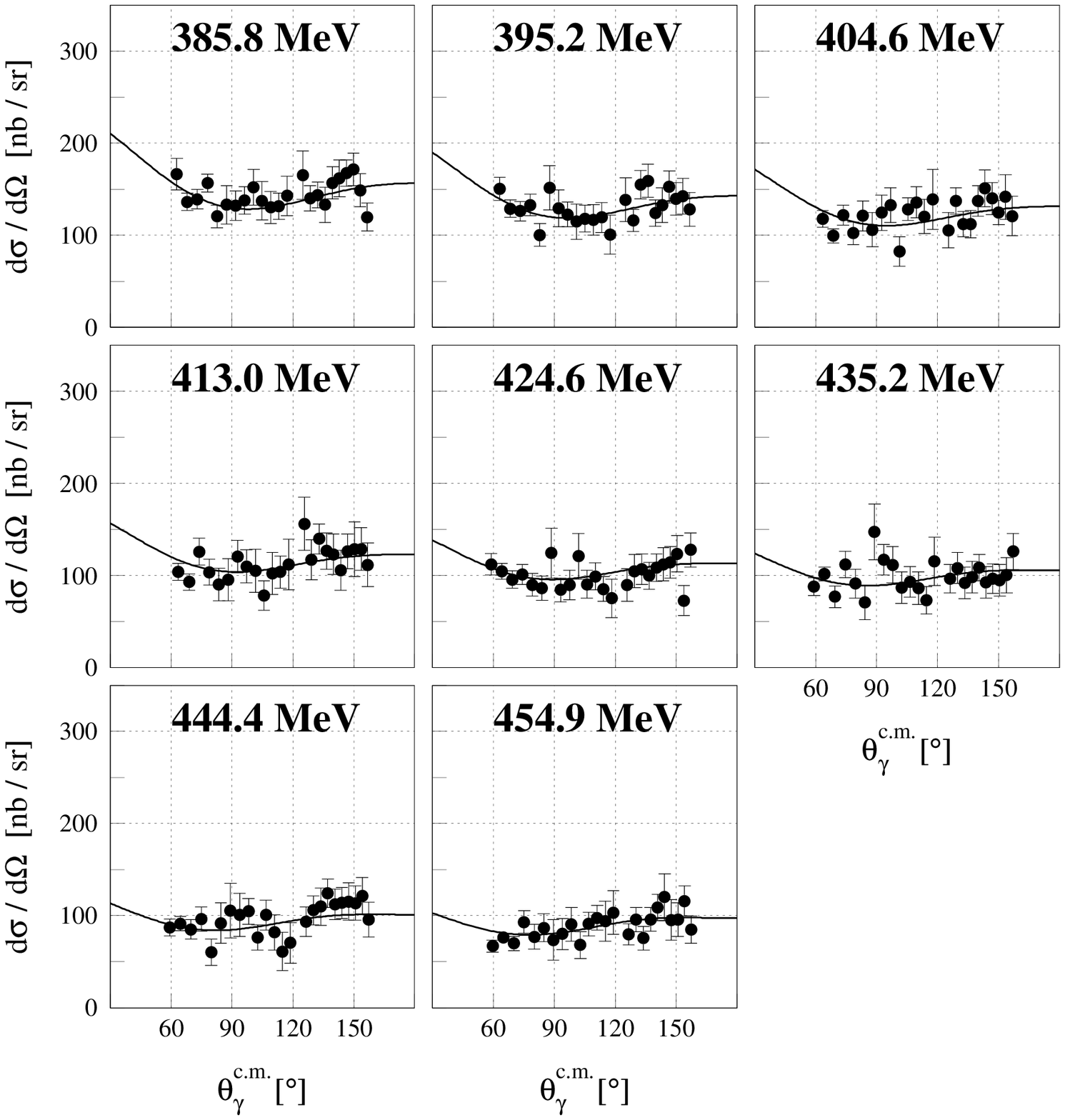}
  \caption{Same as Fig.~\ref{fig:fig13}.}
  \label{fig:fig14}
\end{figure*}
\begin{figure*}
  \centering \includegraphics[width=0.8\textwidth]{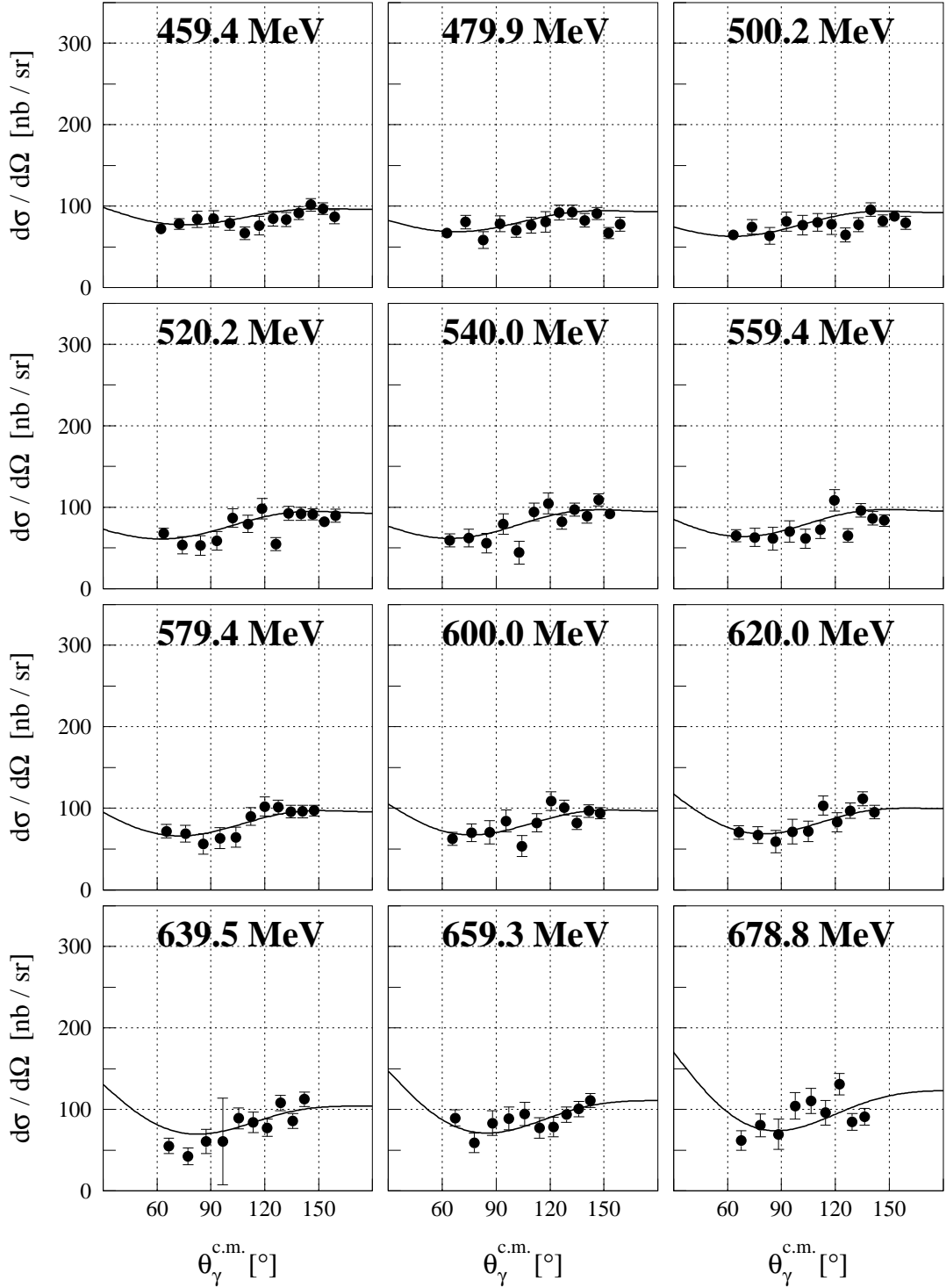}
  \caption{Angular distributions of the differential cross section in the
           c.m.-system as obtained with the LARA arrangement using the
           ``one-dimensional'' analysis in combination with an out-of-plane
           subtraction. Solid line: calculation within the
           dispersion relation approach \protect\cite{lvov97} 
           using the 'best-fit'
           parameters of Eqs.~(\ref{freevariation}.}
  \label{fig:fig15}
\end{figure*}
\begin{figure*}
  \centering \includegraphics[width=0.8\textwidth]{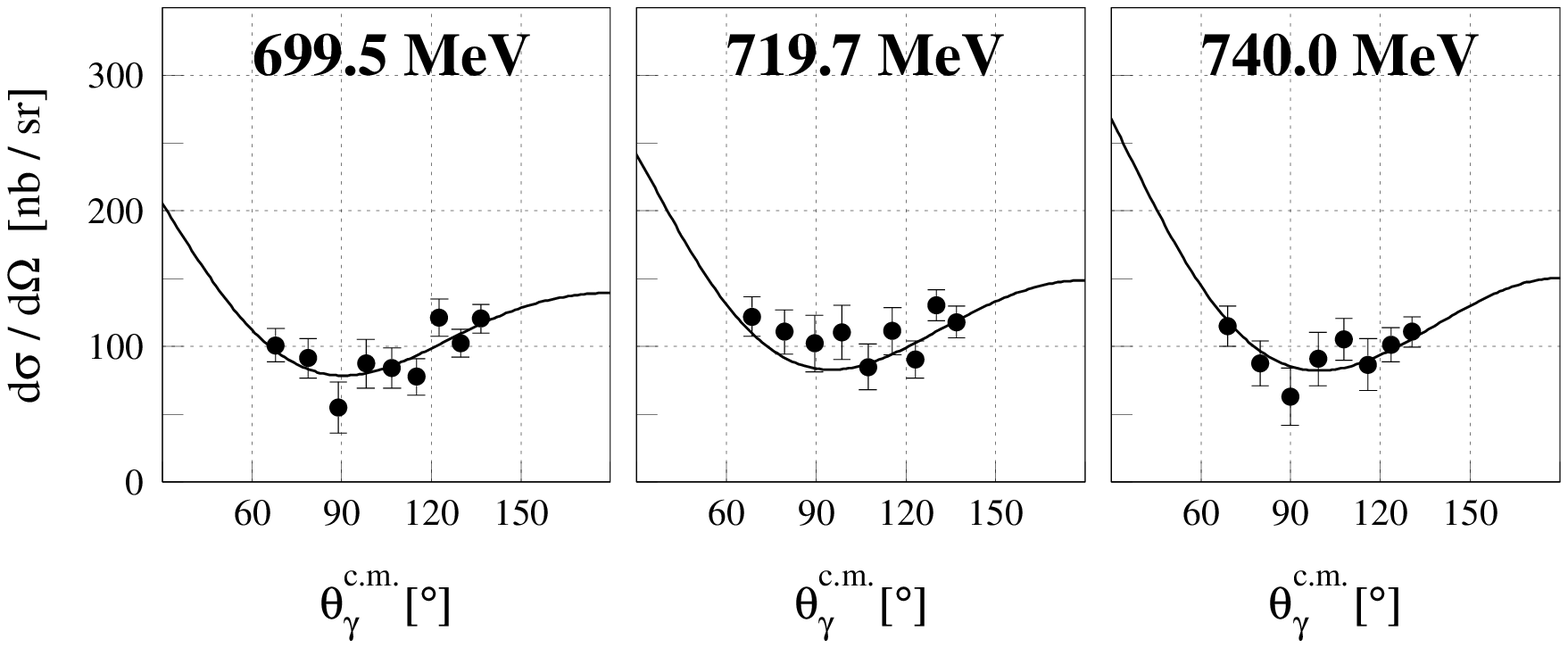}
  \caption{Same as Fig.~\ref{fig:fig15}.}
  \label{fig:fig16}
\end{figure*}
\begin{figure*}
  \centering \includegraphics[width=0.8\textwidth]{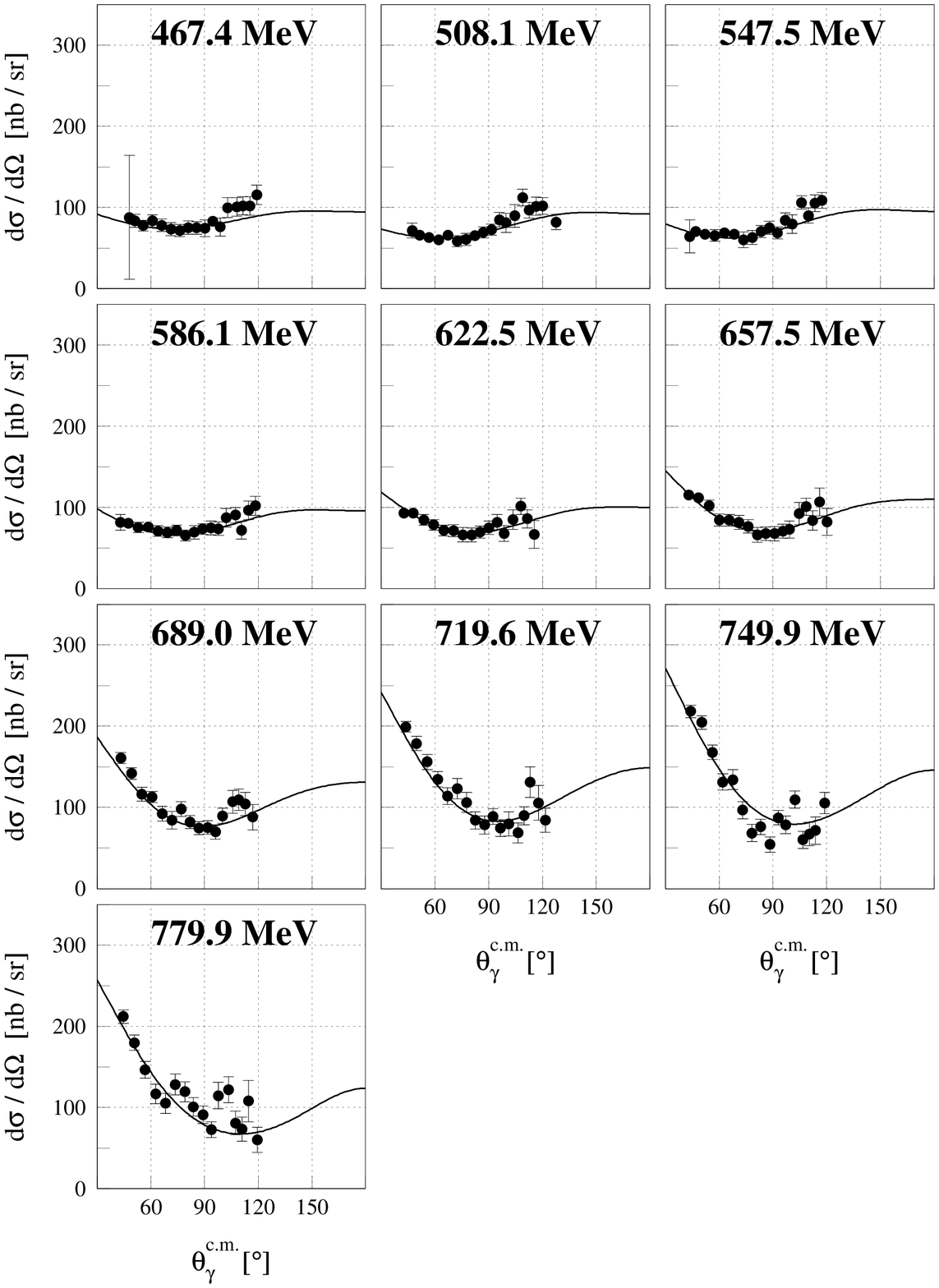}
  \caption{Angular distributions of the differential cross section in the
           c.m.-system as obtained with the LARA arrangement using the
           ``two-dimensional'' analysis. Solid line: calculation within the
           dispersion relation approach \protect\cite{lvov97} 
           using the 'best-fit'
           parameters of Eqs.~(\ref{freevariation}).}
  \label{fig:fig17}
\end{figure*}

\clearpage

\clearpage

%%%%%%%%%%%%%%%%%%%%%%%%%%%%%%%%%%%%%%%%%%%%%%%%%%%%%%%%%%%%%%%%%%%
%%%
%%%     B I B L I O G R A P H Y
%%%

\end{document}